# Noncovalent Functionalization of Black Phosphorus

Gonzalo Abellán,[a] Vicent Lloret, [a] Udo Mundloch, [a] Mario Marcia,[a] Christian Neiss,[b] Andreas Görling,[b] Maria Varela,[c] Frank Hauke[a] and Andreas Hirsch*[a]

[a] Dr. G. Abellán, V. Lloret, U. Mundloch, M. Marcia, Dr. F. Hauke, Prof. A. Hirsch, Chair of Organic Chemistry II and Joint Institute of Advanced Materials and Processes (ZMP), Friedrich-Alexander-Universität Erlangen-Nürnberg (FAU), Henkestr. 42, 91054 Erlangen, Germany. E-mail: *andreas.hirsch@fau.de
[b] Dr. M. Varela, Universidad Complutense de Madrid, Instituto Pluridisciplinar & Facultad de CC. Físicas, Madrid 28040, Spain.
[c] Dr. C. Neiss, Prof. A. Görling, Lehrstuhl für Theoretische Chemie, Friedrich-Alexander-Universität Erlangen-Nürnberg (FAU), Egerlandstrasse 3, 91058 Erlangen, Germany.

**Abstract:** The first non-covalent functionalization of black phosphorus (BP) is presented. The treatment of BP with electron- withdrawing 7,7,8,8-tetracyano-*p*-quinodimethane (TCNQ) leads to electron transfer from BP to the organic dopant. On the other hand, the non-covalent interaction of BP with perylene diimide is mainly due to van-der-Waals interactions but also leads to a considerable stabilization of the BP flakes against oxygen degradation.

BP has recently emerged as a very interesting new two- dimensional material.[1] The individual layers of BP exhibit a honeycomb structure differing to that of graphene because of a marked puckering of the $sp^3$ P-atoms[2,3] Owing to its intrinsic direct bandgap being a good trade-off between charge carrier mobility and current on/off ratios and because of its unusual in- plane anisotropy, BP exhibits a tremendous potential in both electronics and optoelectronics.[4-9]

Few-layer nanosheets (flakes) of BP can be obtained by mechanical exfoliation of the bulk crystals as well as through solvent exfoliation. However, so far practical applications are restricted because of the instability of BP with respect to ambient oxygen and moisture.[10-14] Concepts for chemical stabilization are urgently desired. Encapsulation methods have been proposed to preserve its intrinsic properties, but the chemical control over the reactivity still remains an open challenge.[3] Recently, in collaboration with the group of Coleman, we reported the liquid exfoliation of solvent-stabilized few-layer BP flakes. During this procedure the stability at ambient conditions increased from *ca*. 1h for mechanically cleaved BP to *ca*. 200 h for 1-cyclohexyl-2-pyrrolidone (CHP)-exfoliated few-layer BP. First appealing applications beyond electronics as gas sensors, saturable absorbers and reinforcing fillers for nanocomposites could be demonstrated.[15] The chemistry of BP, however, remains almost unexplored.[16] So far only single flake chemistry with diazonium salts and $TiL_4$ complexes have been reported.[23,24]

We describe now for the first time non-covalent functionalization of bulk BP with TCNQ and perylene bisimides (PDI) demonstrating pronounced charge-transfer and van-der-Waals-interactions. It was possible to isolate and completely characterize functionalized few-layer BP. The experimental results were supported by quantum mechanical calculations.

The chemical approach with which we propose to develop the non-covalent functionalization involves the combination of pristine BP crystals with TCNQ – an extremely versatile building block for charge-transfer compounds[19] – under inert conditions using an argon-filled glovebox ($O_2$ < 0.1 ppm and $H_2O$ < 0.1 ppm; see methods) and mild magnetic stirring (Scheme 1). The appropriate selection of the solvent has a tremendous influence on the process. We initially anticipated that high-boiling point solvents like NMP or CHP were expected to form tightly packed solvation shells adjacent to BP surfaces acting as a barrier to oxygen/water. However, a screening study with different solvents showed that surprisingly TCNQ is able to oxidize some of these systems creating charge-transfer complexes (see Supplementary Methods and Fig. S1). As depicted in Fig. 1a,b, TCNQ exhibits characteristic absorption spectra for the neutral (TCNQ), the aromatized anion radical ($TCNQ^{\cdot-}$) and the dianion ($TCNQ^{2-}$).[20] The generation of the $TCNQ^{2-}$ requires an exhaustive $2e^{-}$ reduction (bulk electrolysis) and working under rigorous exclusion of $O_2$.[20] When the study was performed under extremely inert conditions as we did, several charge- transfer steps between the solvents and the TCNQ took place (Fig. S1). Indeed, the spontaneous formation of the monoanion $TCNQ^{\cdot-}$ takes place for NMP or CHP, whose spectra reveal the presence of the aromatic form even after exposing the solutions to air. It is most likely that the corresponding radical-cation of the pyrrolidone is formed during this process. Therefore, a suitable selection of an inert solvent was required for developing the redox non-covalent functionalization of BP. To this end, THF turned out to be the most appropriate since and it does not react with BP.

For the preparation of the targeted non-covalent hybrids between BP and TCNQ, a mixture of *ca*. 5 mg of grinded BP in 3mL of a $1E^{-5}$ M TCNQ solution in THF was vigorously stirred during a minimum of 3 days in the glovebox, carefully controlling the $O_2$ content during all the working period. Indeed, the high quality of the solvent is also reflected in the absence of water absorption peaks in the nIR spectra at around 1430 nm (Fig. S1). Subsequently, a dark colloidal suspension exhibiting the characteristic Faraday-Tyndall effect was obtained. The first clue of the success was offered by UV/Vis spectroscopy, showing the exclusive formation of a species with a similar UV/Vis spectrum as the TCNQ dianion (hereinafter BP–TCNQ).

Fig. 1c shows a titration experiment testing different TCNQ concentrations ranging from $1E^{-5}$ to $1E^{-2}$M, keeping BP constant. The amount of $TCNQ^{2-}$ decreased with the concentration, exhibiting a marked shift in the bands for values higher than $1E^{-3}$M. Free TCNQ only appeared at concentrations $\geq 1E^{-3}$M, therefore, the optimal working range was $1E^{-5}$–$1E^{-4}$M, which we then used for the sample preparation and further characterization. Additional evidence of the formation of the charge-transfer complex was provided by ATR-FTIR.

In the ATR-FTIR spectrum of pristine THF suspensions of BP, TCNQ, and the functionalized BP (Fig. 1d), one can see that the parent TCNQ exhibits the characteristic vibration frequencies of the cyano group at *ca*. 2224 $cm^{-1}$.

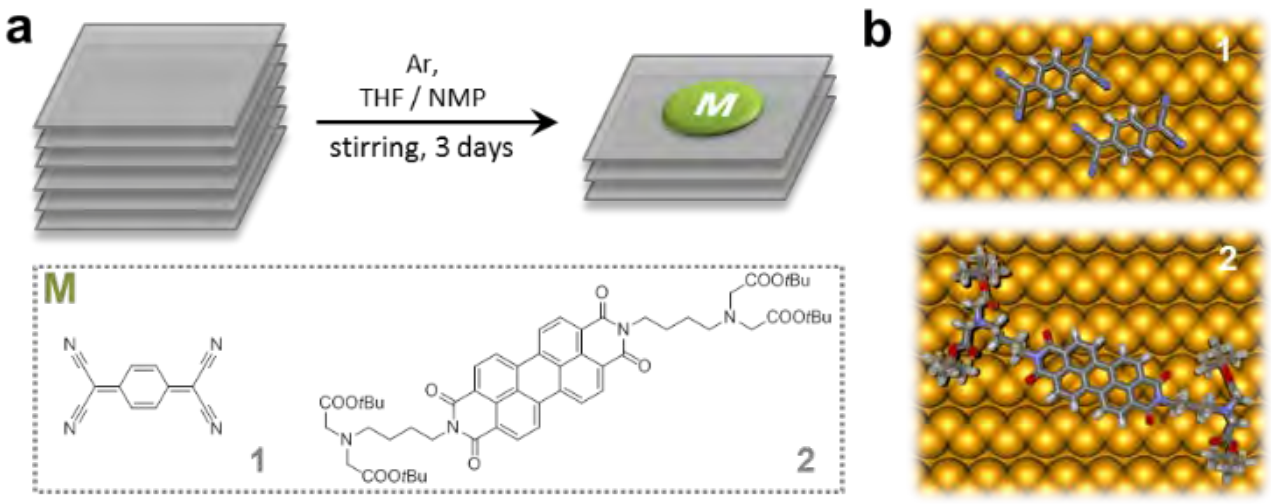

**Scheme 1.** Functionalization scheme. (**a**) Representation of the functionalization of BP without sonication using a strong electron-withdrawing acceptor, TCNQ, and bulk black phosphorous, leading to the formation of charge transfer compounds consisting on few-layers BP decorated with TCNQ Second functionalization approach is based on a novel EDTA-ligand containing an integral perylene bisimide (PDI) core. (**b**) Schematic representation of TCNQ (top) and PDI molecules (bottom) adsorbed on BP.

In stark contrast, the BP–TCNQ complex exhibits a doublet for the cyano group at 2199 and 2168 $cm^{-1}$, indicative of the formation of the $TCNQ^{\cdot -}$, as well as a shoulder at around 2131 $cm^{-1}$, related to the $TCNQ^{2-}$ formation (Fig. S2). These observations are in accord with previously reported samples of $TCNQ^{2-}$-containing coordination polymers,[21–23] and point towards the formation of an anionic TCNQ species on BP under inert conditions, as previously predicted by theoretical calculations.[24–26]

Compound BP–TCNQ was also investigated by scanning Raman microscopy (SRM). Typical results are depicted in Fig. 1e-g. We have selected different areas using an optical microscope, and measured the corresponding spectra with an excitation wavelength of 532 nm (Fig. S3). All the spectra showed the characteristic modes of BP, labelled $A_g^1$, $B_{2g}$ and $A_g^2$. We performed a statistical Raman mapping measuring a selected flake in Fig. 1e. We therefore analysed and plotted the $A_g^1/A_g^2$ intensity ratio showing a well-defined image of the micrometric flake (with an overall thickness of *ca*. 20 nm subsequently determined by AFM, and corresponding to less than 35 layers), highlighting that there were no areas with $A_g^1/A_g^2 < 0.4$ confirming the basal planes to be slightly oxidized.[11] However, significant differences can be observed if compared with CHP solvent-stabilized few-layers BP, in which $A_g^1/A_g^2 > 0.5$, indicative of the redox reaction with the TCNQ in addition to oxygen degradation.[15] Control experiments under inert conditions revealed very similar results (average $A_g^1/A_g^2$ ratio around 0.6), showing a slight oxidation in the basal plane (see Fig. S4). Consequently, we searched for the characteristic bands of anionic TCNQ. Well-defined $\nu 3$ and $\nu 4$ modes at *ca*. 1580 and 1329 $cm^{-1}$, respectively, were encountered at the flake edges, whereas only broad contributions were measured in the basal planes, pointing towards a preferential accumulation of TCNQ at the edges. (Fig. S4b).[27]

Despite the contributions from residual solvent and adhesion effects, the apparent AFM thickness provides useful information about the exfoliation degree of BP–TCNQ. We have relocated under the AFM the same flake previously studied by SRM (Fig. S3). The inset in Fig. 1e shows a typical thickness of *ca*. 20 nm with lateral dimensions of several microns (see Fig. S5).

1 3 2

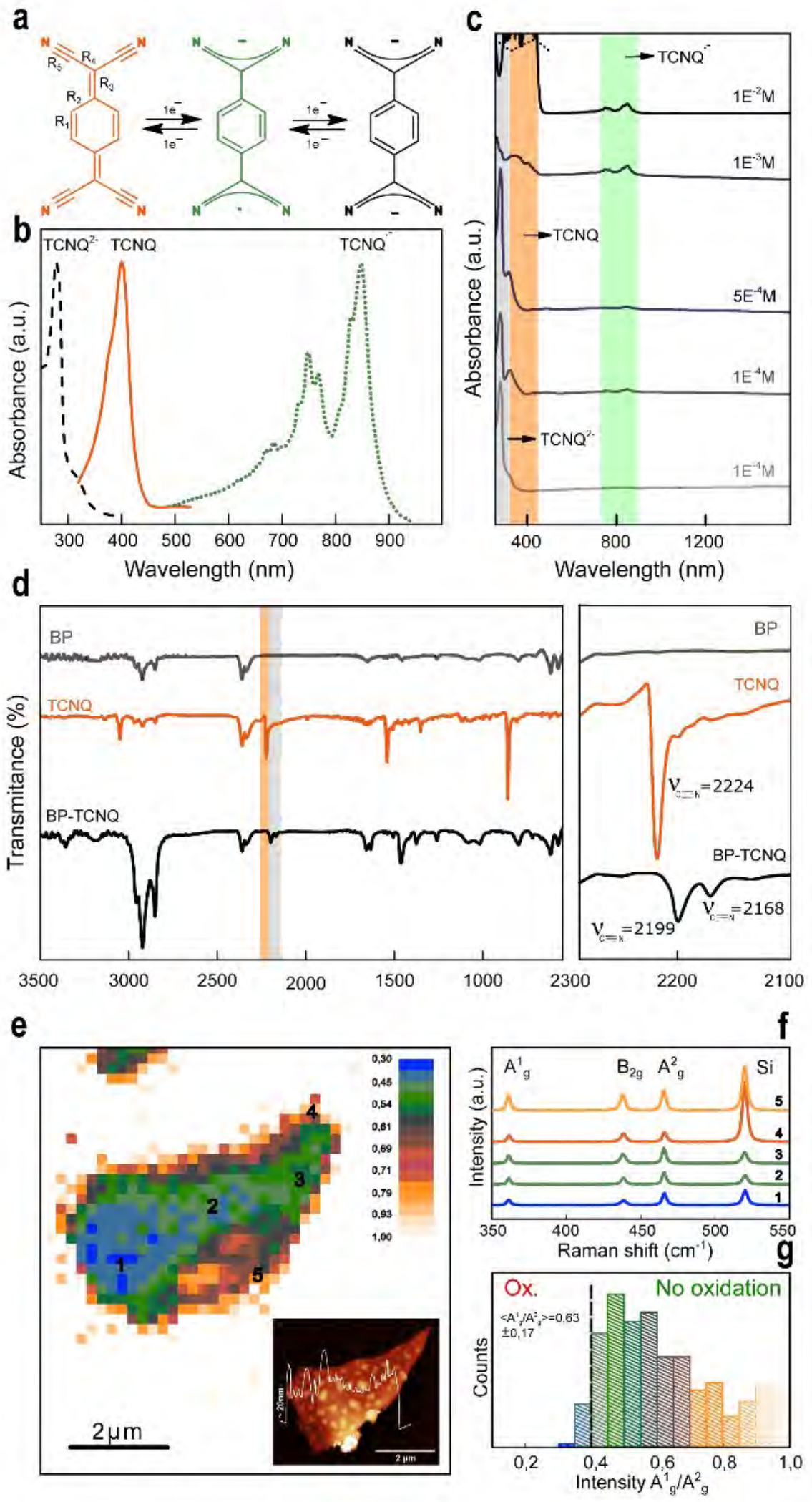

**Figure 1.** Synthesis and characterization of TCNQ functionalized BP nanosheets. (**a**) Molecular structure of TCNQ, TCNQ·⁻, and TCNQ²⁻. The TCNQ molecule exhibits a quinonoid structure, with a high single–double bond length alternancy, where R and R are much shorter than R .[28] (**b**) Electronic absorption spectra fingerprints of the different TCNQ species in THF solution.[20] TCNQ·⁻ absorption spectrum obtained by exhaustive exclusion of $O_2$ in NMP, and TCNQ²⁻ obtained in the presence of BP (see Methods). (**c**) Titration experiment with BP highlighting the exclusive formation of BP-TCNQ below 1E⁻³ M concentrations. The typical absorption bands of TCNQ·⁻ and TCNQ²⁻ are indicated as bars. (**d**) ATR-FTIR spectra of pristine BP (grey), TCNQ (orange), and BP–TCNQ (black). (**e**) SRM of a BP–TCNQ ($10^{-5}$ M) flake showing the corresponding $A_{1g}/A_{2g}$-band ratio mapping. (**f**) Mean Raman spectra (excitation 532 nm) of the areas indicated in (**e**). (**g**) Histogram of the $A_{1g}/A_{2g}$ intensity ratio. The inset in (**e**) represents an AFM image of the same flake studied by SRM.

The characteristic degradation protuberances usually ascribed to the formation of $H_3PO_3$ and/or $H_3PO_4$ were also evident in the topographic image (taken after the SRM analysis was finished, *i.e.* around 24h after its first contact with the oxygen), indicating that the interaction with the TCNQ molecules didn't prevent the degradation of the flakes (see Fig. S4). Thinner flakes can be encountered, but they also exhibit smaller lateral dimensions (< 1 micron), precluding its use in further statistical Raman analysis (see Fig. S5).

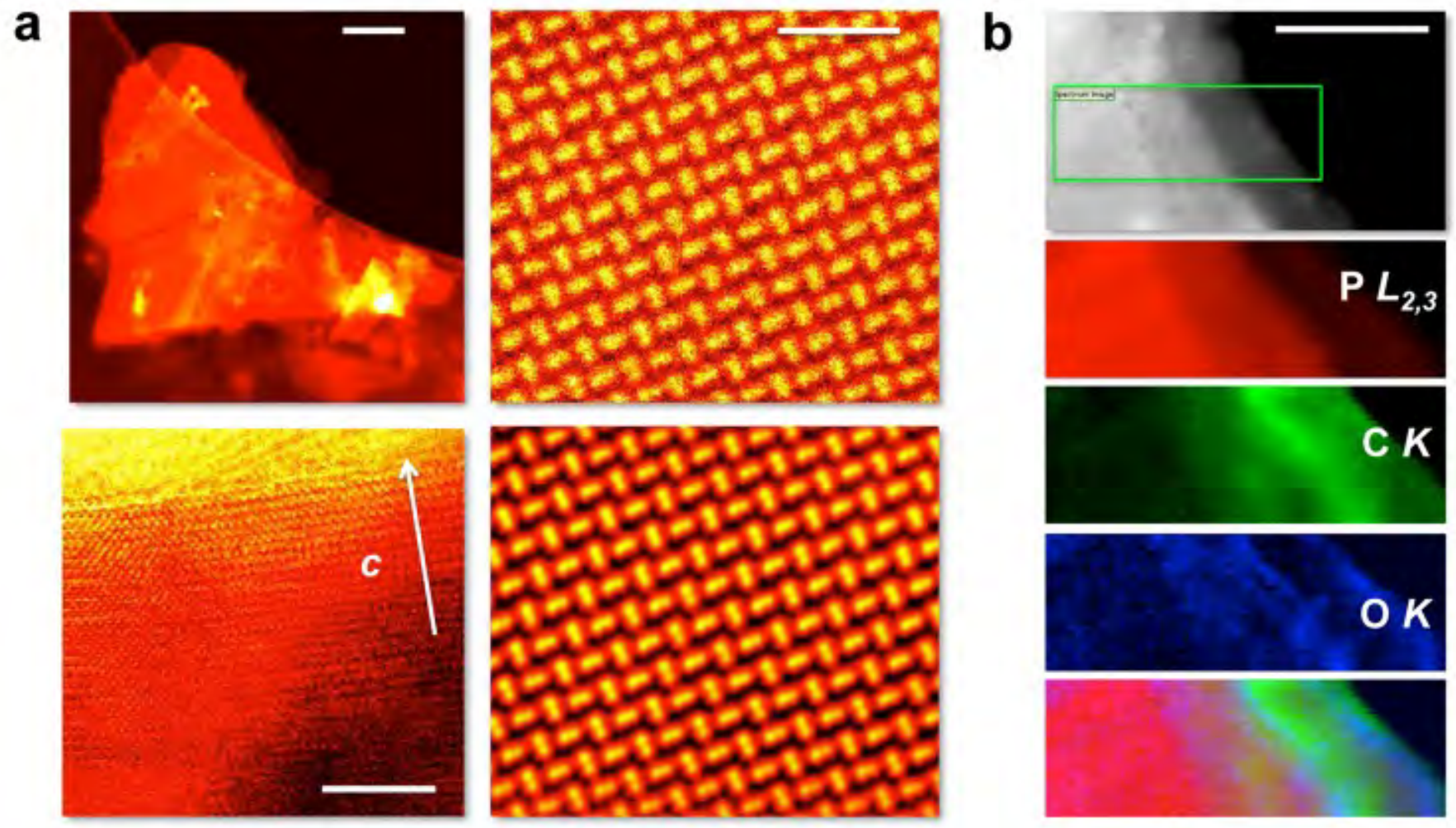

**Figure 2:** STEM–EELS analysis of BP-TCNQ nanosheets. (**a**) STEM images of a flake (BP–TCNQ $10^{-5}$ M): low magnification HAADF (top left, scale bar is 200 nm), ABF image of a flake with the c axis perpendicular to the electron beam (bottom left), along with atomic resolution HAADF, down the [101] axis (right panels). The top panel exhibits the raw data, while the bottom panel displays a Fourier filtered image. All images in false color, with the scale bars for the atomic resolution images representing 1 nm. (**b**) Compositional maps obtained on the edge of a flake. The top panel displays the area where an EEL spectrum image was acquired, marked with a green rectangle. The bottom panels show the P, C, and O maps, along with an RGB overlay following the same color scheme. The scale bar represents 50 nm.

Herein, our strategy represents a milder alternative to our previously described sonication procedure and yields larger flakes (inset Fig. 1e and Fig. S5). For direct microscopic evidence of the non-covalent functionalization of the BP sheets we analyzed both, the pristine BP and BP–TCNQ samples, by real space techniques sensitive to both chemistry and structure such as atomic resolution aberration-corrected STEM-EELS at 80 kV. Crystal sizes were in the range of microns, as depicted in the low magnification high angle annular dark field (HAADF) image of Fig. 2a (left). An atomic resolution HAADF STEM image of the BP–TCNQ sample down the [101] crystallographic direction is shown on the right panel (both raw and Fourier filtered data), acquired with the electron beam perpendicular to the platelet plane. The sample is highly crystalline showing the characteristic puckered structure that appears intact over wide regions, of the order of hundreds of nanometers. The lattice shows high uniformity with the presence of very few defects or dislocations. Depending on the zone axis used, the lattice can exhibit a “dumbbell” type configuration. The observed average thickness (estimated from low loss EELS measurements) revealed values in the 8–20 nm range in good accordance with the AFM measurements. Cross- sectional views of the crystals such as the annular bright field (ABF) image in the bottom panel of Fig. 2a show parallel fringes corresponding to interlayer distances of 0.5–0.6 nm, which is consistent with previous observations.[29] In order to have an additional evidence of the TCNQ functionalization, the chemical composition of the flake was also analyzed by STEM–EELS. Fig. 2a displays the study of a flake (*ca*. 10 nm overall thick, corresponding to *ca*. 15–18 layers) suspended over a hole in order to avoid any contribution from the carbon-coated copper grid. Fig. 3b shows the chemical maps obtained from the $L_{2,3}$– edges of P, the C *K* edge, and the O *K* edge, with onsets near 132, 284 and 532 eV, respectively. These maps, which are more sensitive to the surface molecule coating than HAADF imaging, have been produced by subtracting the background using a power-law fit and then integrating the remaining signal below the edges of interest, using 25 eV wide windows. Carbon signals can be observed on all areas, pointing to a homogeneous anionic TCNQ surface coverage. A significant accumulation of carbon can be detected on surface steps and flake edges, in agreement with the SRM results. On the other hand, the oxygen mapping reveals a residual contribution mainly located on the edges of the flakes, probably arising from the grid transfer into the TEM holder (< 30 sec.). Moreover, we excluded the presence of THF residues in the basal plane of the flakes by EELS analysis. Control experiments on pristine BP in THF also revealed the absence of carbon, discarding the presence of solvent residues. Additionally, EELS mapping showed the predominant presence of oxygen on the edges and terraces exhibiting highly degraded rims, with amorphous edges and the presence of holes. These results confirmed the predicted preferential edge degradation of BP (Fig. S6).[15]

To provide clearer experimental evidence of the non-covalent functionalization in BP, we took advantage of the chemical versatility of the family of perylene diimide (PDI)-based molecules. Indeed, our group has used these molecules for the non-covalent functionalization of carbon nanotubes, graphene or MoS2, exhibiting a strong absorption due to their large aromatic core that can get attached to the 2D layers *via* van-der-Waals interactions.[30–32] The structure of the selected molecule is depicted in Scheme 1, we used a protected EDTA-PDI molecule in order to avoid any preferential interaction of the carboxylic moieties (see Fig. S7 for synthetic information).[33] Control experiments revealed that NMP can generate highly colored charge-transfer complexes with the PDI molecules, when working under strictly inert conditions, which exhibits a dramatic sensitivity against oxygen (see Fig. 3b and Fig. S8). This striking result will be fully developed and analyzed in a separate work. The first signature of the PDI functionalization can be substantiated by fluorescence spectroscopy (Fig. 3c). Upon excitation at 455 nm, the PDI emission exhibited a dramatic quenching of the fluorescence of ca. 66 % in the presence of BP, in excellent accordance with previous studies for graphene functionalization (Fig. 3c).[31,34]. Although perylene diimides exhibit a very strong fluorescence – which usually prevents their investigation by Raman spectroscopy – we were able to relocate with AFM appropriate flakes and record a Raman spectrum of the BP–PDI comprising both the features of BP and PDI, when excited at 532 nm (Fig. 3d and Fig. S9). In presence of BP a complete quenching of the fluorescence takes place, exhibiting the characteristic signals of the PDI spectrum (Fig. 3e,f and Fig. S10).

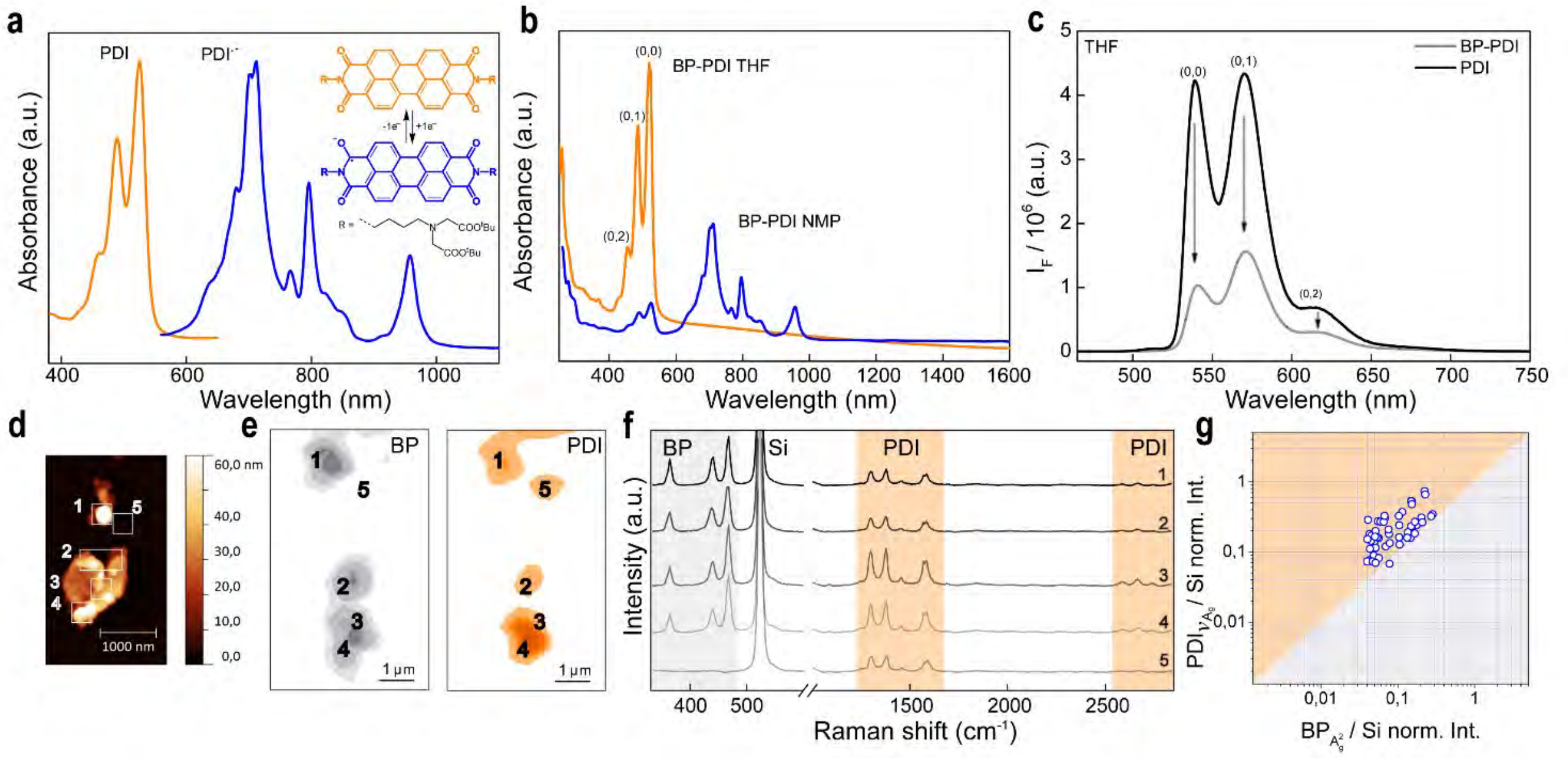

**Figure 3**: Synthesis and characterization of PDI functionalized BP nanosheets. (**a**) Molecular structure of the PDI and $PDI^{\bullet-}$. Electronic absorption spectra fingerprints of the different PDI species. (**b**) UV-Vis spectra of BP-PDI in THF and NMP. (**c**) Fluorescence emission spectra of PDI and BP–PDI in THF ($c_{PDI}$ = $10^{-4}$ M, dotted line); $\lambda_{exc}$ = 455 nm. (**d**) Representative AFM image and its corresponding Raman $A^1_g$ (**e**, $\lambda_{exc}$ = 532 nm) and $\nu_{Ag}$ intensities mappings (**f**) of the same BP–PDI flakes. The numbers denote the areas where the Raman spectra shown in (**f**) were taken. (**g**) Plot of the $A^1_g$ versus $\nu_{Ag}$ normalized intensities.

The statistical analysis of Raman spectra comparing the Si- normalized intensities of the $\nu_{Ag}$ mode of PDI with the $A^1_g$ mode of BP revealed a comparable intensity ratio (Fig. 4e-g and Fig. S11–13 for additional AFM and Raman analyses). Strikingly, when we analysed the $A_g^1/A_g^2$ ratio after 2 days under ambient conditions, we observed an average value of *ca*. 0.79 similar to that shown by freshly exfoliated flakes and remarkably higher than that exhibited by BP–TCNQ, indicative of non-oxidized samples (Fig. S14). Additionally, we measured a substrate after 6 months stored in a glove box, observing that there is no degradation (Fig. S15). We also analysed the thermal stability of these flakes after air exposure by statistical temperature-dependent Raman spectroscopy (Fig. S16). The STEM-EELS analysis, summarized in Fig. 4, further confirmed the presence of the organic moiety on the surface of the exfoliated crystals, exhibiting flakes within a thickness range of 15–50 nm (estimated from low loss EELS). A low magnification image of a flake a few microns in size is displayed in the top panel.

Flakes were crystalline, as shown by atomic resolution images. Fig. 4a (middle) exhibits an intermediate magnification ABF image exhibiting long-range crystalline order. The inset displays a high magnification HAADF image down the [101] axis. Fig. 4a (bottom panel) shows a flake with the [001] axis perpendicular to the electron beam. The interatomic spacing is of the order of 5.5–6.0 Å, and the surface appears coated with an organic layer a few nm thick. A study of the chemical composition by EELS is summarized in Fig. 4b, where P (red), C (green), and O (blue) maps are shown from top to bottom, along with an RGB superposition of the three of them. This analysis reveals a complete coverage of the flakes with C (the PDI moiety), being more pronounced in the edges. Very little oxygen signal was detected in this sample (within the experimental noise). This finding suggests that the PDI protective shell showed a thickness of a few nm (3–5 nm visible in the bottom panel of Fig. 4a), which acts as an encapsulation reagent preventing the environmental degradation of the BP.

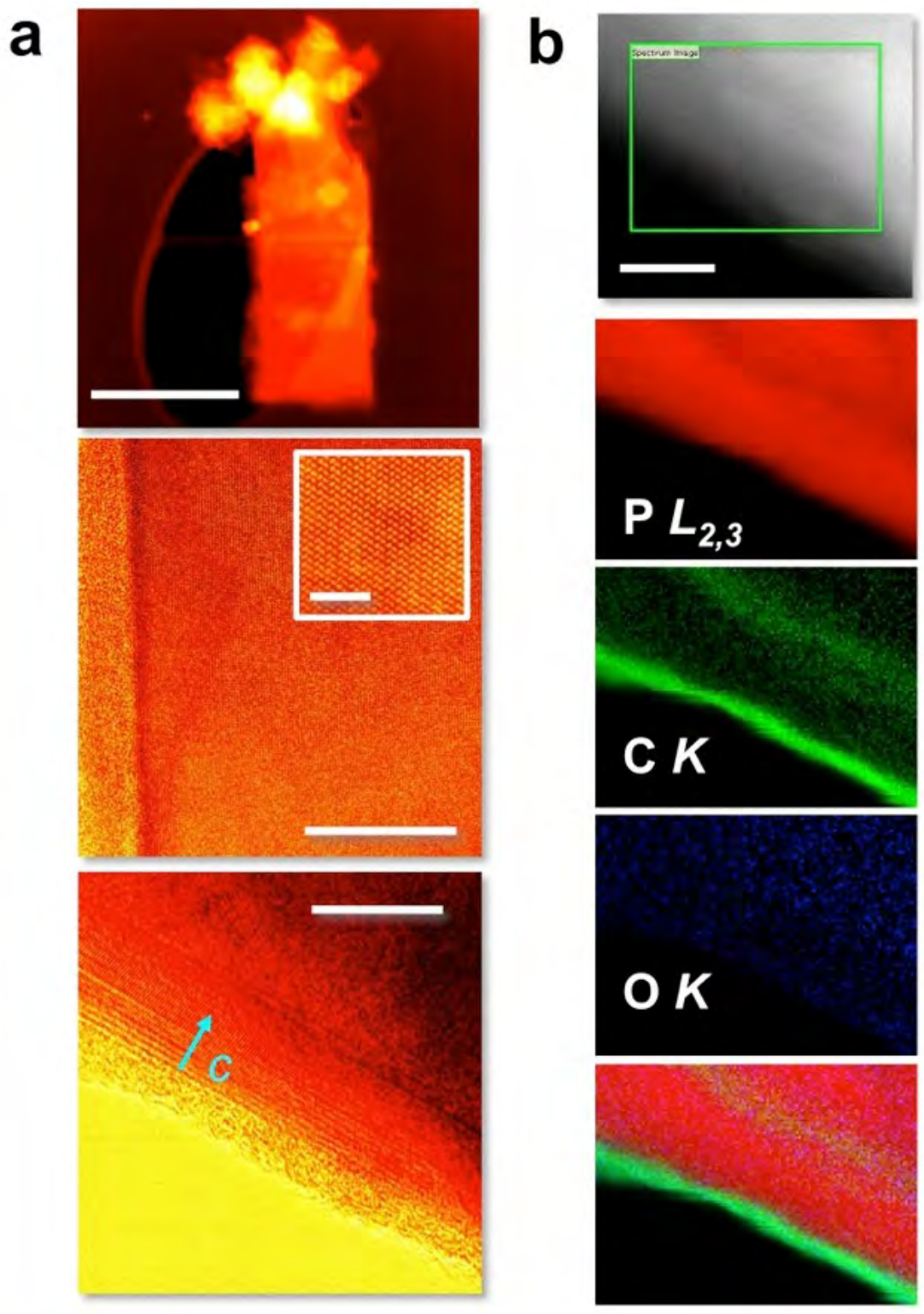

**Figure 4.** STEM–EELS analysis of BP-PDI nanosheets. (**a**). STEM images of a PDI coated flake. From top to bottom: low magnification HAADF image (top), ABF image of a flake with the [101] axis perpendicular to the electron beam (middle), along with atomic resolution HAADF, down the [101] direction (in the inset), ABF image of the edge of a flake (bottom). All images in false color. The scale bars, from top to bottom, correspond to 500 nm, 20 nm (inset is 2 nm) and 10 nm, respectively. (**b**) Compositional maps, the top panel shows an HAADF image of the area where an EEL spectrum image was acquired, marked with a green rectangle. The scale bar represents 10 nm. The bottom panels show the P $L_{2,3}$, C *K*, and O *K* maps from this area, along with an RGB overlay following the same color scheme.

Finally, we systematically investigated the supramolecular interaction of TCNQ and PDI on BP by DFT calculations. In the case of TCNQ the oxidation of the BP flakes was confirmed.

Significantly, increasing the number of layers of BP causes increasing of the amount of electron transfer. Due to the lower reduction potential of PDI only weak charge transfer was calculated (see Fig. S17 for detailed information).

In conclusion, we have presented here for the first time a concept for the non-covalent organic functionalization of black phosphorus (BP). The wet-chemical treatment of BP with the electron poor and polarizable polycyclic aromatics, namely, 7,7,8,8-tetracyano-p-quinodimethane (TCNQ) molecules and a tailor-made perylene diimide leads to the formation of stable hybrids were the organic components cover and shield the surface of thin few layer BP flakes. Experimental and computational studies revealed the nature of the strong non-covalent interaction between these molecules and BP, provoking its exfoliation into few-layers. In the case of TCNQ electron transfer from the BP to the organic component takes place. The positive charge on the BP surface is stabilized by the layers underneath which is nicely supported by quantum mechanical calculations. As a consequence, even the formation of the $TCNQ^{2-}$ dianion is supported, which can be easily monitored by absorption spectroscopy. The functionalization with perylene diimide improves dramatically the resistance of the flakes against oxygen degradation. We expect that with this initiation of fundamental organic chemistry of black phosphorus, access to unprecedented 2D-materials is provided. The inherent modulation of the physical properties of black phosphorus will allow for targeting applications in electronics, optoelectronics, energy storage, sensors or fillers for composite reinforcement, to name a few.

**Acknowledgements**

The research leading to these results has received funding from the European Union Seventh Framework Programme under grant agreement n°604391 Graphene Flagship. The authors thank the Deutsche Forschungsgemeinschaft (DFG-SFB 953 “Synthetic Carbon Allotropes”, Projects A1 and C2), the Interdisciplinary Center for Molecular Materials (ICMM), and the Graduate School Molecular Science (GSMS) for financial support. Electron microscopy was performed at the Centro Nacional de Microscopia Electrónica (UCM) sponsored by the ERC Starting Investigator Award STEMOX#239739 and Fundación BBVA. G.A. acknowledges the EU for a Marie Curie Fellowship (FP7/2013-IEF-627386).

[1] A. C. Ferrari, F. Bonaccorso, V. Fal'ko, K. S. Novoselov, S. Roche, P. Bøggild, S. Borini, F. H. L. Koppens, V. Palermo, N. Pugno, et al., Nanoscale 2015, 7, 4598–4810.

[2] X. Ling, H. Wang, S. Huang, F. Xia, M. S. Dresselhaus, Proc. Natl. Acad. Sci. 2015, 112, 4523–4530.

[3] A. Castellanos-Gomez, J. Phys. Chem. Lett. 2015, 6, 4280–4291.

[4] J. D. Wood, S. A. Wells, D. Jariwala, K.-S. Chen, E. Cho, V. K. Sangwan, X. Liu, L. J. Lauhon, T. J. Marks, M. C. Hersam, Nano Lett. 2014, 14, 6964–6970.

[5] A. Castellanos-Gomez, L. Vicarelli, E. Prada, J. O. Island, K. L. Narasimha-Acharya, S. I. Blanter, D. J. Groenendijk, M. Buscema, G. A.Steele, J. V. Alvarez, et al., 2D Mater. 2014, 1, 025001.

[6] M. Buscema, D. J. Groenendijk, G. A. Steele, H. S. J. van der Zant, A. Castellanos-Gomez, Nat. Commun. 2014, 5, 4651.

[7] J. Qiao, X. Kong, Z.-X. Hu, F. Yang, W. Ji, Nat. Commun. 2014, 5, 4475.

[8] F. Xia, H. Wang, Y. Jia, Nat. Commun. 2014, 5, 4458.

[9] L. Li, Y. Yu, G. J. Ye, Q. Ge, X. Ou, H. Wu, D. Feng, X. H. Chen, Y. Zhang, Nat. Nanotechnol. 2014, 9, 372–377.

[10] J. Kang, J. D. Wood, S. A. Wells, J.-H. Lee, X. Liu, K.-S. Chen, M. C. Hersam, ACS Nano 2015, 9, 3596–3604.

[11] A. Favron, E. Gaufrès, F. Fossard, A.-L. Phaneuf-L'Heureux, N. Y.-W. Tang, P. L. Lévesque, A. Loiseau, R. Leonelli, S. Francoeur, R. Martel, Nat. Mater. 2015, 14, 826–832.

[12] J. O. Island, G. A. Steele, H. S. J. van der Zant, A. Castellanos- Gomez, 2D Mater. 2015, 2, 011002.

[13] P. Yasaei, B. Kumar, T. Foroozan, C. Wang, M. Asadi, D. Tuschel, J. E. Indacochea, R. F. Klie, A. Salehi-Khojin, Adv. Mater. 2015, 27, 1887–1892.

[14] A. H. Woomer, T. W. Farnsworth, J. Hu, R. A. Wells, C. L. Donley, S. C. Warren, ACS Nano 2015, 9, 8869–8884.

[15] D. Hanlon, C. Backes, E. Doherty, C. S. Cucinotta, N. C. Berner, C. Boland, K. Lee, A. Harvey, P. Lynch, Z. Gholamvand, et al., Nat. Commun. 2015, 6, 8563.

[16] C. R. Ryder, J. D. Wood, S. A. Wells, M. C. Hersam, ACS Nano 2016, DOI 10.1021/acsnano.6b01091.

[17] C. R. Ryder, J. D. Wood, S. A. Wells, Y. Yang, D. Jariwala, T. J. Marks, G. C. Schatz, M. C. Hersam, Nat. Chem. 2016, advance online publication, DOI 10.1038/nchem.2505.

[18] Y. Zhao, H. Wang, H. Huang, Q. Xiao, Y. Xu, Z. Guo, H. Xie, J. Shao, Z. Sun, W. Han, et al., Angew. Chem. Int. Ed. 2016, 55, 5003–5007.

[19] T.-C. Tseng, C. Urban, Y. Wang, R. Otero, S. L. Tait, M. Alcamí, D. Écija, M. Trelka, J. M. Gallego, N. Lin, et al., Nat. Chem. 2010, 2, 374–379.

[20] M. R. Suchanski, R. P. Van Duyne, J. Am. Chem. Soc. 1976, 98, 250– 252.

[21] L. Ballester, A. M. Gil, A. Gutiérrez, M. F. Perpiñán, M. T. Azcondo, A.

E. Sánchez, E. Coronado, C. J. Gómez-García, Inorg. Chem. 2000, 39, 2837–2842.

[22] M. R. Saber, A. V. Prosvirin, B. F. Abrahams, R. W. Elliott, R. Robson, K. R. Dunbar, Chem. – Eur. J. 2014, 20, 7593–7597.

[23] X. Zhang, M. R. Saber, A. P. Prosvirin, J. H. Reibenspies, L. Sun, M. Ballesteros-Rivas, H. Zhao, K. R. Dunbar, Inorg Chem Front 2015, 2, 904–911.

[24] R. Zhang, B. Li, J. Yang, J. Phys. Chem. C 2015, 119, 2871–2878.

[25] Y. Jing, Q. Tang, P. He, Z. Zhou, P. Shen, Nanotechnology 2015, 26, 095201.

[26] Y. He, F. Xia, Z. Shao, J. Zhao, J. Jie, J. Phys. Chem. Lett. 2015, 6, 4701–4710.

[27] E. I. Kamitsos, W. M. R. Jr, J. Chem. Phys. 1983, 79, 5808–5819.

[28] B. Milián, R. Pou-Amérigo, R. Viruela, E. Ortí, J. Mol. Struct. THEOCHEM 2004, 709, 97–102.

[29] R. J. Wu, M. Topsakal, T. Low, M. C. Robbins, N. Haratipour, J. S. Jeong, R. M. Wentzcovitch, S. J. Koester, K. A. Mkhoyan, J. Vac. Sci. Technol. A 2015, 33, 060604.

[30] C. Backes, C. D. Schmidt, K. Rosenlehner, F. Hauke, J. N. Coleman, A. Hirsch, Adv. Mater. 2010, 22, 788–802.

[31] N. V. Kozhemyakina, J. M. Englert, G. Yang, E. Spiecker, C. D. Schmidt, F. Hauke, A. Hirsch, Adv. Mater. 2010, 22, 5483–5487.

[32] C. Wirtz, T. Hallam, C. P. Cullen, N. C. Berner, M. O'Brien, M. Marcia, A. Hirsch, G. S. Duesberg, Chem. Commun. 2015, 51, 16553–16556.

[33] M. Marcia, P. Singh, F. Hauke, M. Maggini, A. Hirsch, Org. Biomol. Chem. 2014, 12, 7045.

[34] J. M. Englert, J. Röhrl, C. D. Schmidt, R. Graupner, M. Hundhausen, F. Hauke, A. Hirsch, Adv. Mater. 2009, 21, 4265–4269.

[35] X. Ling, W. Fang, Y.-H. Lee, P. T. Araujo, X. Zhang, J. F. Rodriguez- Nieva, Y. Lin, J. Zhang, J. Kong, M. S. Dresselhaus, Nano Lett. 2014, 14, 3033–3040.

# Supporting Information

## Noncovalent Functionalization of Black Phosphorus

Gonzalo Abellán,[1] Vicent Lloret,[1] Udo Mundloch,[1] Mario Marcia,[1] Christian Neiss,[3] Andreas Görling,[3] Maria Varela,[2] Frank Hauke[1] and Andreas Hirsch[1 *]

[1] *Chair of Organic Chemistry II and Joint Institute of Advanced Materials and Processes (ZMP), Friedrich-Alexander-Universität Erlangen-Nürnberg (FAU), Henkestr. 42, 91054 Erlangen, Germany.*

[2] *Universidad Complutense de Madrid, Instituto Pluridisciplinar & Facultad de CC. Físicas, Madrid 28040, Spain.*

[3] *Lehrstuhl für Theoretische Chemie, Friedrich-Alexander-Universität Erlangen-Nürnberg (FAU), Egerlandstrasse 3, 91058 Erlangen, Germany.*

**andreas.hirsch@fau.de*

**Experimental Section.**

*Materials*

BP crystals were purchased from Smart Elements (purity 99.998%); THF, DME, DMSO, benzonitrile, pentane, CHP, NMP, IPA, TCNQ, and all other chemicals were purchased from Sigma Aldrich.

*Functionalization procedure in the glove box*

The different solvents (THF, DME, NMP, CHP, DMSO, benzonitrile, and pentane; Sigma-Aldrich) were dried by allowing them to stand over a desiccant (4 Å molecular sieves from Sigma-Aldrich pre-dried at 300 °C under argon atmosphere for 72 h) under argon for 96 h. A coulometric Karl Fischer titrator was used to determine the water content. Residual traces of oxygen were removed by pump freeze treatment (five iterative steps) before introducing the solvents into the argon-filled Labmaster sp Glovebox (MBraun), equipped with a gas filter to remove solvents and an argon cooling systems, with an oxygen content <0.1 ppm and a water content <0.1 ppm.

Upon introduction to the Glovebox, the samples were stored in dark bottles until the experiment were completed. The mother solutions of TCNQ and PDI were prepared *in situ* with an initial concentration of $1E^{-2}$ M in THF or NMP, which was afterwards diluted to the desired concentration. BP crystal was ground inside the Glovebox and used as a crushed powder. In a typical procedure an excess of 5 mg of BP were added to 3 mL of the final solutions.[1,2] The samples were magnetically stirred *in situ* at 1100 rpm for a total of 3–5 days. After stirring the samples were subjected to centrifugation at 1000 g for 1 minute *in situ* to remove the non-exfoliated particles.

For the absorption spectroscopy (Lambda 1050, see below) or fluorescence spectroscopy (Horiba Scientific Fluorolog-3, see below) the samples were disposed in sealed vials. After the measurements, the samples were discarded. For the entire study the samples were stored under conditions with water and oxygen contents of < 0.1 ppm.

*Karl Fisher titrations of solvents*

To determine the water content of solvents we used the "Karl Fischer titration" a well-known water analysis method (Metrohm USA Inc., http://www.metrohmusa.com/Service/FAQ/Titration/Coulo.html). The Coulometric Karl

Fischer reaction involves adding a known mass of sample to the reaction vessel containing a base, alcohol, and iodine. A current change in the electrode is noted which relates to a chemical change occurring in the vessel.

*Characterization*

Optical extinction and absorbance was measured on a Perkin Elmer Lambda 1050 spectrometer in extinction, in quartz cuvettes with a path length of 0.4 cm. Fluorescence was acquired on a Horiba Scientific Fluorolog-3 system equipped with 450 W Xe halogen lamp, double monochromator in excitation (grating 600 lines/mm blazed at 500 nm) and emission (grating 1200 lines/mm blazed at 500 nm) and a PMT detector using quartz cuvettes with a path length of 1.0 cm. Spectra were obtained at room temperature. ATR-FTIR spectra were recorded on a Bruker Tensor FTIR spectrometer equipped with ZnSe.

Real space studies of the samples were carried out by aberration corrected scanning transmission electron microscopy (STEM) and electron energy-loss spectroscopy (EELS) in a JEOL JEM-ARM200CF electron microscope equipped with a spherical aberration corrector, a cold field emission gun and a Gatan Quantum EEL spectrometer, operated at 80 kV. The measurements were performed at the Centro Nacional de Microscopia Electrónica (UCM), Madrid, Spain. The suspended BP samples were dropped onto lacey carbon coated copper TEM grids under strictly inert conditions. The samples were then stored in vacuum overnight. The images were recorded using a 80 kV acceleration voltage using both a high field annular dark field detector (HAADF) and annular bright field imaging (ABF). The suspended samples were dropped onto lacey carbon coated copper TEM grids as described above. For the PDI samples, the grids were submitted to air ion cleaner prior to be measured (EC-52000IC Ion Cleaner). The images were recorded using a 80 kV acceleration voltage using a high field annular dark field detector (HAADF) and low-pass bright field imaging.

Atomic force microscopy (AFM) was carried out using a Solver Pro scanning probe microscope (NT-MDT Co) in tapping mode and high accuracy HA_NC Etalon tips, as well as NSG10_DLC with 1–3 nm tip radius (both from NT MDT Co). The samples were prepared by drop casting a solution of the phosphorous samples on heated Si/$SiO_2$ wafers with an oxide layer of 300 nm, at 120°C under inert conditions. Typical image sizes were 3–10 µm at scan rates of 0.4-0.6 Hz. To facilitate deposition, the BP–PDI samples in NMP were transferred to IPA by centrifugation at 15 krpm (22,640 g, rotor 1195-A), removal of NMP supernatant and reagitation of the sediment in IPA.

Raman spectroscopy on individual flakes was performed using a Horiba Jobin Yvon LabRAM Aramis confocal Raman spectrometer equipped with a microscope and an automated XYZ–table (excitation wavelength 532 nm) with a laser spot size of ∼1 μm (Olympus LMPlanFl 100x, NA 0.80). The influence of the step size and the oxidation degree in the scanning Raman microscopy was determined by testing micromechanically-exfoliated flakes prepared and measured under inert conditions (see Supplementary Fig. 18 for additional information). The functionalized samples were prepared by drop casting a solution of the phosphorous compounds on heated $Si/SiO_2$ wafers with an oxide layer of 300 nm, at 120°C under inert conditions. The incident laser power was kept as low as possible to avoid structural sample damage: 240 μW (532 nm), and the grating was 1800 $g\cdot mm^{-1}$. The spectra were recorded in both inert atmosphere (using a home-made sealed holder) and air under ambient conditions. Relocalization was achieved by using the optical contrast of nanomaterials deposited on opaque bilayered substrates.

*DFT calculations:*

All DFT computations were performed using the projector-augmented plane wave method as implemented in the Vienna ab-initio simulation package[3,4] within the slab model approach. We adopted the exchange-correlation functional according to Perdew, Burke and Ernzerhof (PBE)[5] together with the van-der-correction D3 by Grimme[6] (in connection with Becke-Johnson damping[7]) to account for van-der-Waals interactions. The plane-wave energy cut-off was 450 eV. Geometry optimizations were carried out using the "Gadget" driver by Bučko *et al.*[8] and were considered converged when all residual force components were smaller than 0.01 eV/Å. Vibrational frequencies were computed by finite differences, in these cases the geometry was optimized until all force components were smaller than 0.001 eV/Å. The Brillouin zone was sampled by 3x3x1 Monkhorst-Pack *k*-point sets.[9] For the final energies the Tetrahedron method including Blöchl corrections for interpolation between the *k*-points was employed.[10] For details about the slab-models see supplementary material.

*DFT calculations: slab models*

Single-layer BP was optimized using orthorhombic cells containing four P-atoms. An orthorhombic cell was used to provide better comparability between the results. To avoid interactions between the P-layers (=slabs) a vacuum of ~15 Å was put in-between. A Monkhorst-Pack *k*-point mesh containing 19x19x1 *k*-points was adopted in both cases. The final optimized cell-parameters are 3.292 Å x 4.441 Å for the BP monolayer.

Based on these values a 6x5 supercell for single-layer BP was built, resulting in cells with 19.750 Å x 22.203 Å base area. At the same time the cell dimension perpendicular to the surface ("cell height" in the following) was set to 20 Å. Correspondingly, the *k*-point meshes were reduced to 3x3x1 *k*-points.

The same procedure was repeated for the double-layer BP, yielding an orthorhombic 3.301 Å x 4.311 Å primitive cell. Accordingly, the corresponding 6x5 supercell is 19.806 Å x 21.555 Å. This basal area was also used for intercalation of TCNQ between single-layer and double-layer BP.

For the calculation of the adsorbate structures the cell height was taken 20 Å for adsorption on single surfaces, and 25 Å for adsorption on the BP double-layer and the intercalation between single-layers. The intercalation between BP double-layers was modelled in a cell with a cell height of 35 Å. In all cases a 3x3x1 Monkhorst-Pack *k*-point grid was used. During the geometry optimizations Methfessel-Paxton smearing with σ = 0.2 eV was applied.[11] Final energies were, however, evaluated using the tetrahedron method according to Blöchl.[10]

The isolated TCNQ and PDI molecules were calculated in a 19.750 Å x 22.203 Å x 20 Å cell, i.e. in the optimized supercell of single-layer BP. During the geometry optimizations of the isolated molecules Gaussian smearing with σ = 0.01 eV was applied. As before, final energies were evaluated using the tetrahedron method according to Blöchl.[4]

**Supplementary Figure 1 | Comparative study of the formation of charge-transfer complexes between 7,7,8,8-tetracyano-*p*-quinodimethane (TCNQ) and different solvents.**

The TCNQ solutions (1E$^{-4}$M) in different solvents were prepared *in situ* in an argon-filled glove box under strictly inert conditions (with values of less than 0.1 ppm of $O_2$ and $H_2O$) using: tetrahydrofuran (THF), benzonitrile (BNZ), 1-methyl-2-pyrrolidinone (NMP), 1-cyclohexyl-2-pyrrolidone (CHP), 1,2-dimethoxyethane (DME), dimethyl sulfoxide (DMSO), and pentane as a insoluble reference.

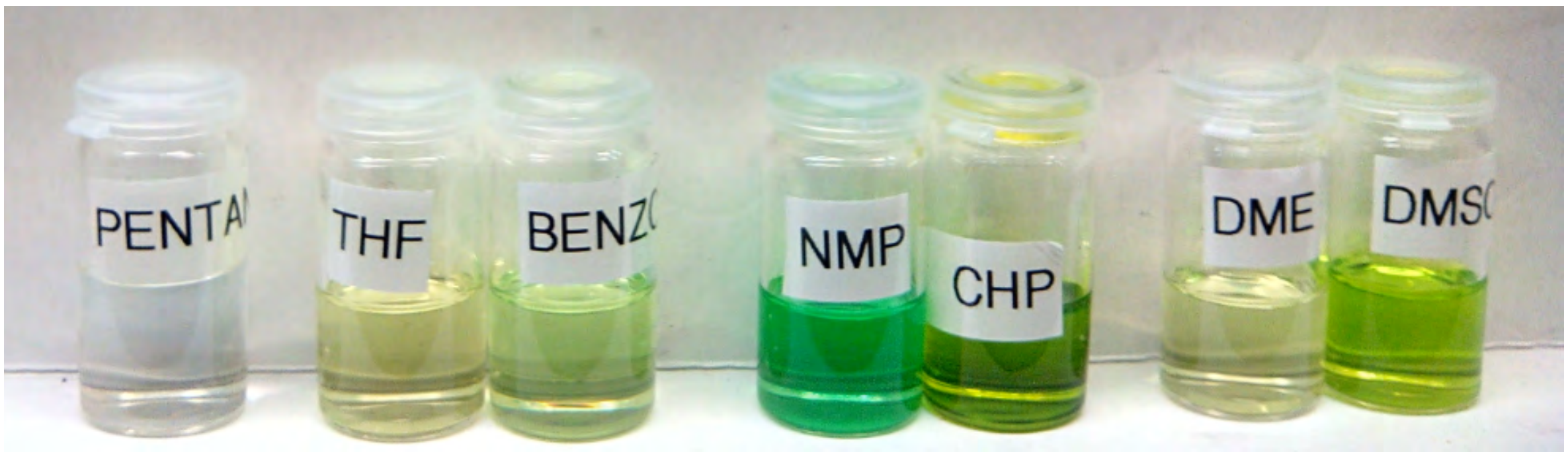

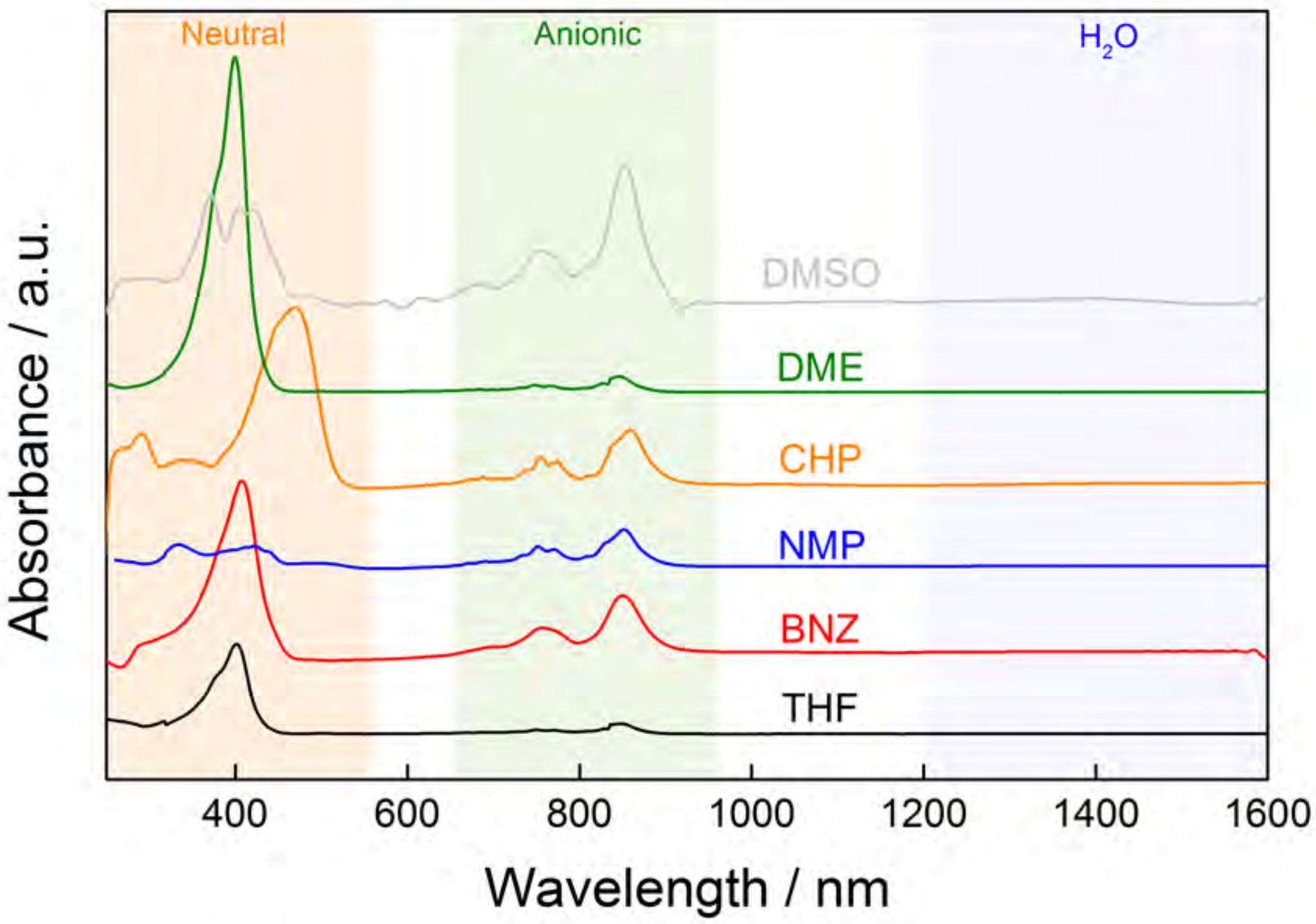

The samples were studied *via* UV/Vis-nIR under argon-sealed cuvettes. The characteristic regions for the neutral and monoanionic species are highlighted in the plot, as well as the absence of $H_2O$ in the 1200–1600 nm region, indicative of the high dryness of the solvents.

**UV/Vis-nIR spectra of NMP-TCNQ and CHP-TCNQ charge-transfer complexes with and without oxygen.**

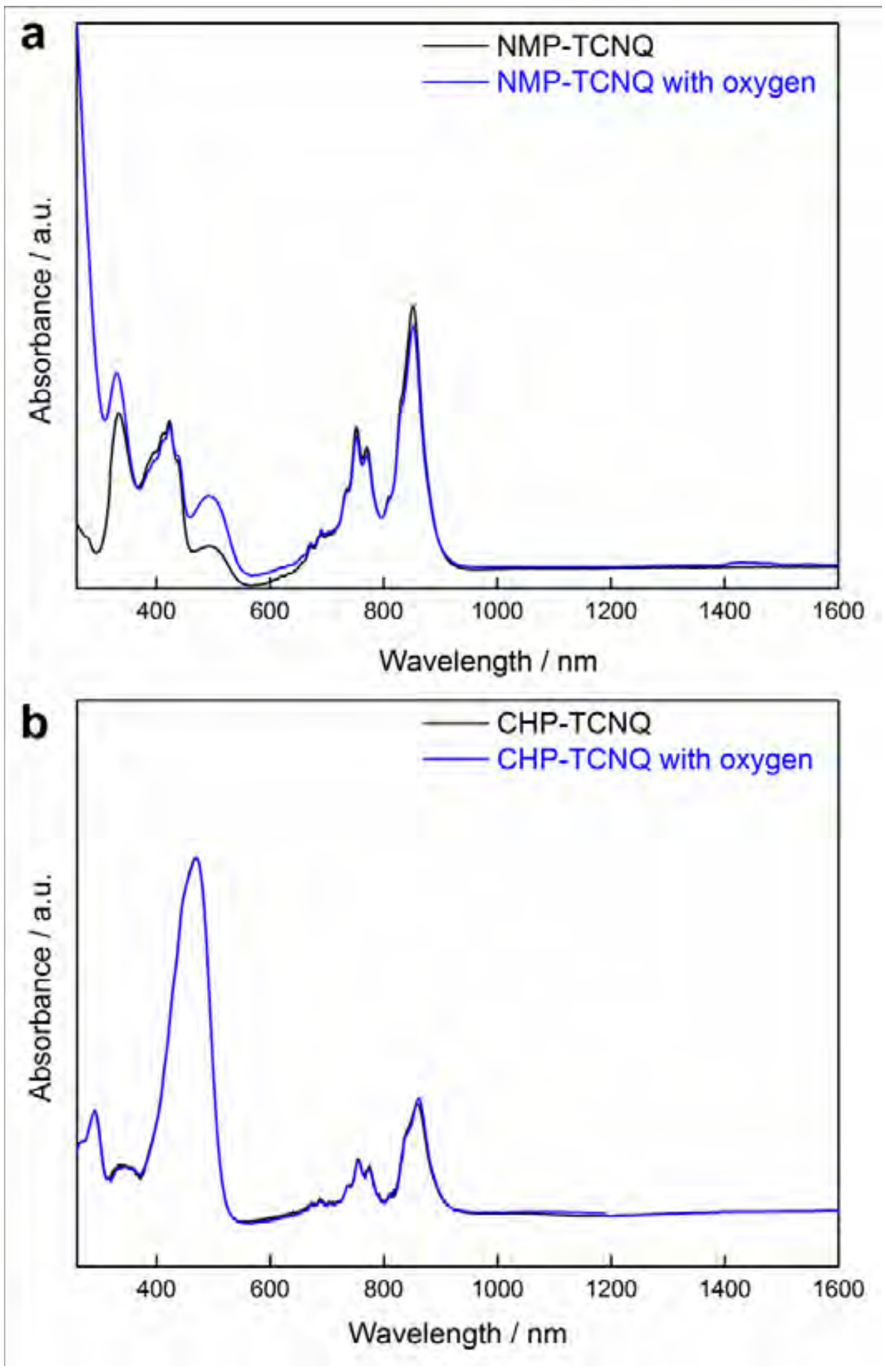

The absorption spectra before and after opening the sealed cuvettes revealed a marginal influence of the oxygen on the absorption bands, highlighting the formation of strong charge-transfer complexes.

It has turned out that the choice of the solvent to be used as the dispersing reagent for wet-chemical functionalization is very crucial. We have discovered that typical solvents such as 1-methyl-2-pyrrolidinone (NMP) or CHP to disperse 1D- and 2D-nanomaterials undergo an intrinsic electron transfer process with TCNQ. Although this discards them for the targeted functionalization procedure of BP this new finding is highly interesting itself.

**Supplementary Figure 2 | ATR-FTIR analysis of BP-TCNQ.**

As a general rule, the $\nu_{(C\equiv N)}$ modes are expected to shift to lower energies when the negative charge on the TCNQ increases.[12] Especially sensitive are the ungerade bands of $\nu_{20}$ and $\nu_{34}$ (which are nearly degenerate and appear at 1506 $cm^{-1}$ in BP-TCNQ; numbering of the bands is as in Ref. [31]), and $\nu_{50}$ (at 802 $cm^{-1}$, with an additional broad band at *ca*. 821 $cm^{-1}$), which are shifted to lower frequencies when the charge on the TCNQ is increased, a fingerprint of the formation of TCNQ anions. Obviously, as the ATR-FTIR spectrum was acquired under atmospheric conditions, the formation of the monoanion also occurs.

**Supplementary Figure 3 | Optical-AFM relocalization.**

The flake size of the exfoliated BP material can reach the micrometer scale, facilitating a fast exploration of a wafer under the optical microscope, to identify the type and position of the BP, as well as hinting at the expected thickness of each flake, due to a thickness dependent contrast and color. Each wafer was transferred under the AFM, which has a less powerful optical microscope attached. The previously found points of interest can thereby easily be located again, in order to find the right position for the cantilever.

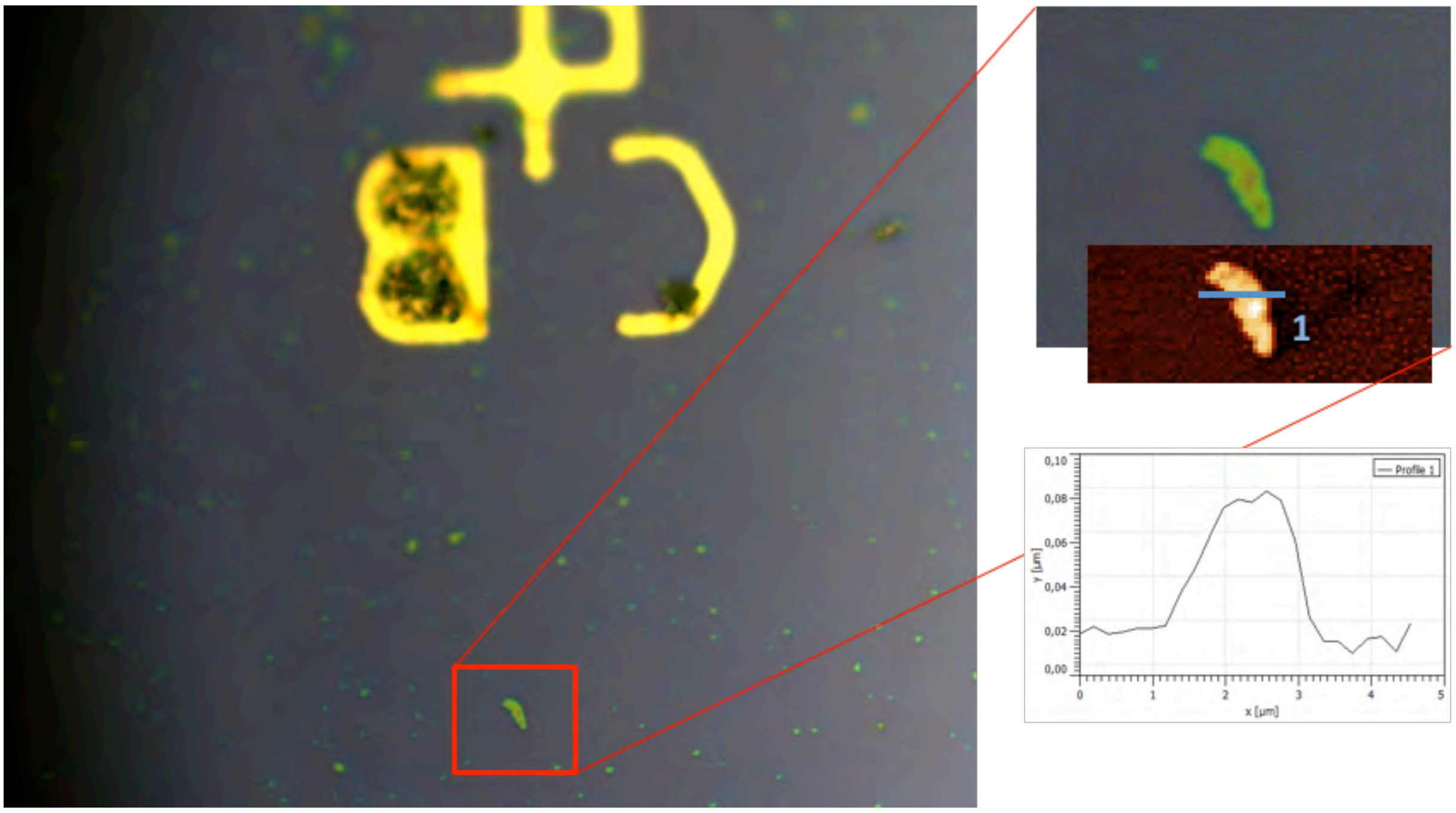

Optical microscopy (100x) of BP flake with large area and contrast hinting at thin structure (left). Magnification of the same flake (top right) and resulting AFM recording (middle right) with profile, showing a thickness of ~60 nm (bottom right).

**Supplementary Figure 4 | Raman analysis of BP–TCNQ under inert conditions.**

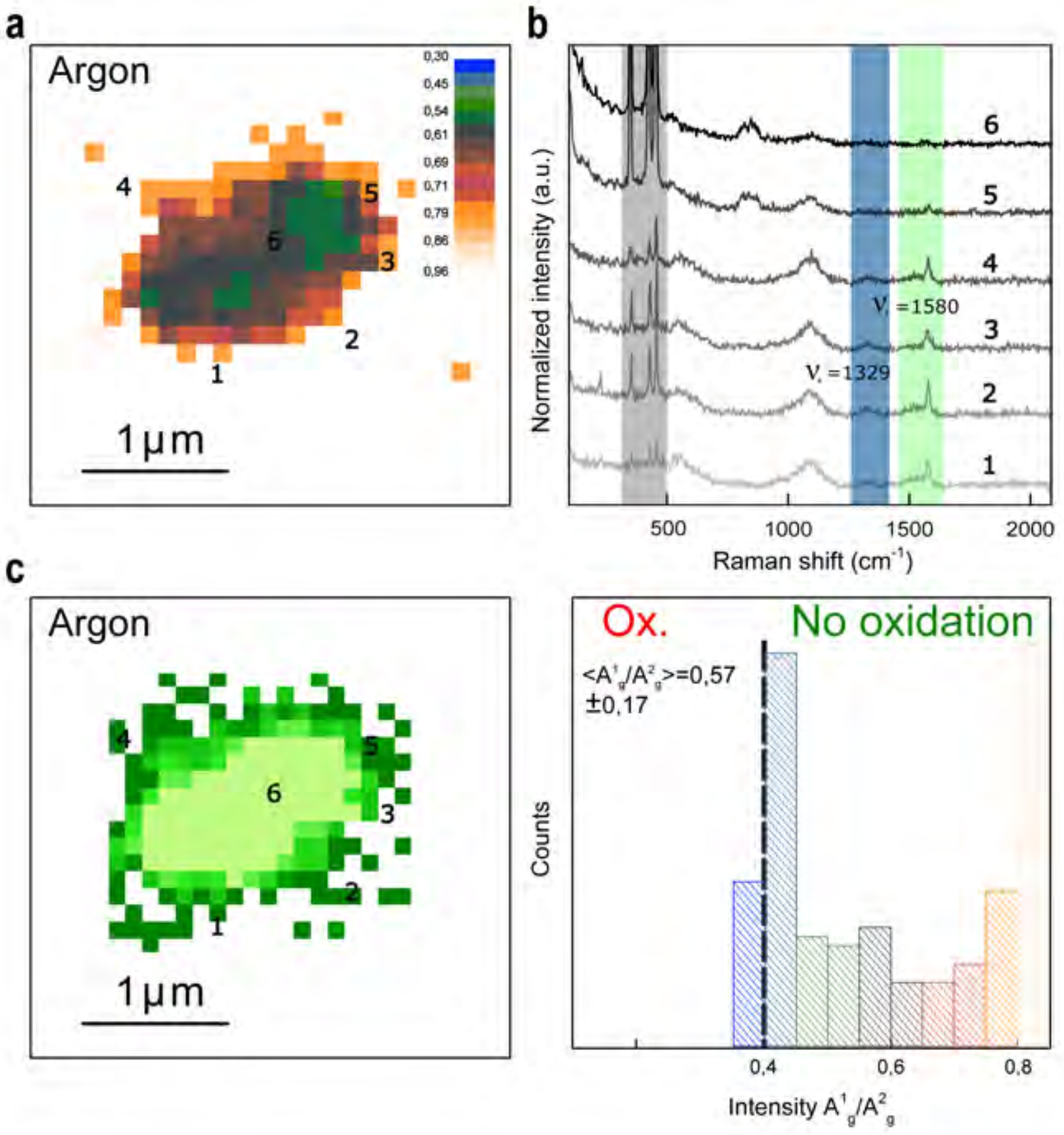

(**a**) Scanning Raman microscopy of a BP–TCNQ flake deposited on a $Si/SiO_2$ substrate with a 300 nm thick oxide layer and measured under inert conditions using an argon-filled sealed holder, showing the $A_{1g}/A_{2g}$-band ratio mapping. (**b**) Raman spectra (excitation 532 nm) of the areas indicated in (**a**) revealing the presence of $TCNQ^{2-}$ characteristic $\nu_3$ and $\nu_4$ modes at *ca*. 1580 and 1329 $cm^{-1}$, respectively.[13–15] It is worth to remark that the TCNQ was hardly detected in the basal plane of the flakes, exhibiting only slight broad contributions, whereas was much more evident in the edges of the flakes, pointing towards a preferential edge accumulation. (**c**) SRM mapping of the $\nu_3$ mode of the TCNQ showing the spatial distribution of the molecule. (**d**) Histogram of the intensity ratio of the $A_{1g}/A_{2g}$ modes obtained from the analysis of the flake showing an average value of 0.57. Interestingly the flake revealed a slight degree of oxidation, probably due to the $TCNQ^{2-}$.[16]

**Supplementary Figure 5 | AFM topographic analysis of a BP–TCNQ sample.**

The solvent assisted exfoliation of BP lead to flakes mostly in the sub 100 nm regime (lateral dimensions).[1,2,17,18] The process could be improved by adding TCNQ without using sonication. While the typical flake thicknesses in solvents alone reached up to 1 µm, the system with TCNQ in THF easily facilitated single, unbroken flakes with 20–30 nm thin but ~6 µm in lateral dimensions.

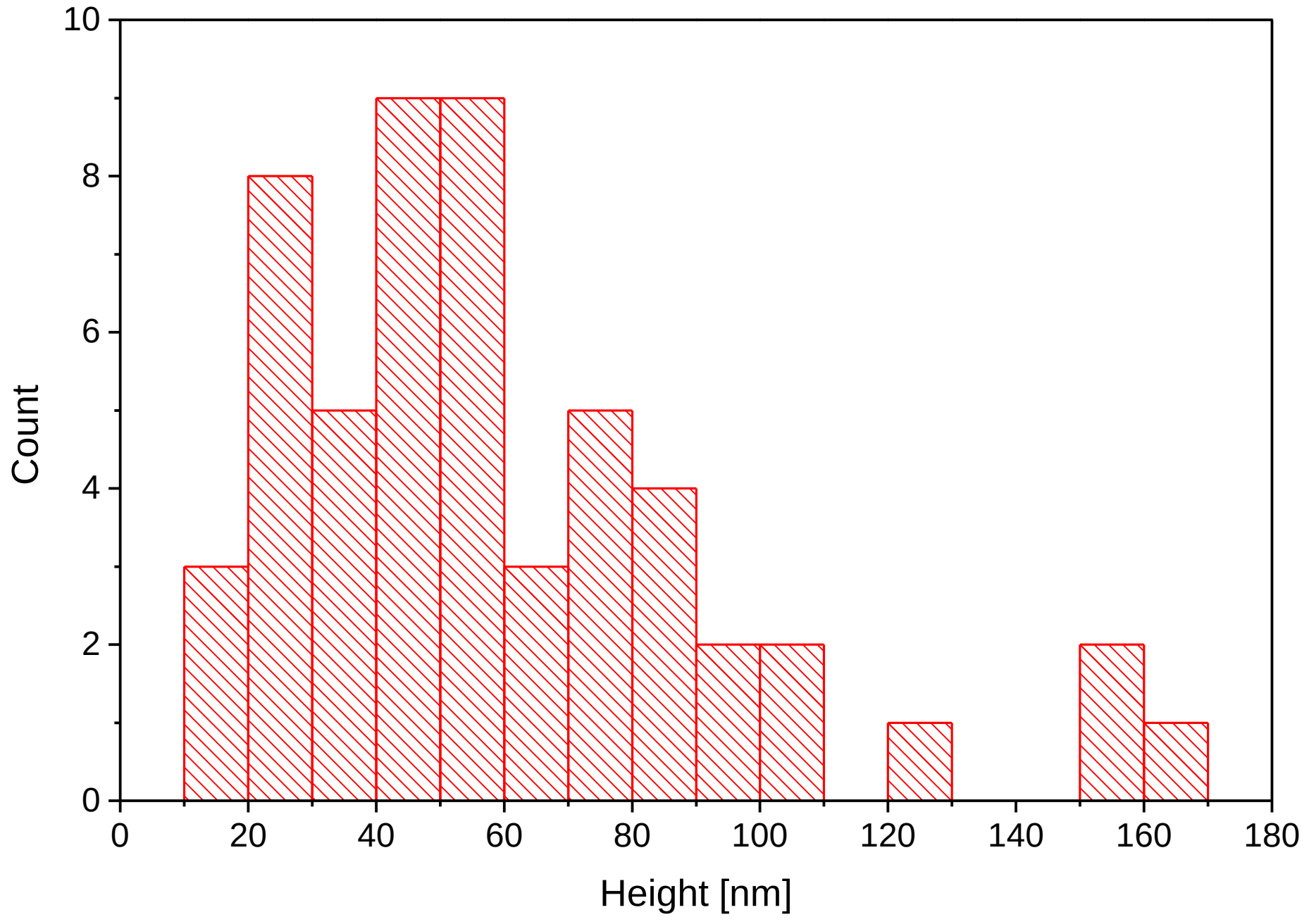

The information on flake thickness was collected in order to compare our different approaches of BP exfoliation. While the difficulty to find measurable material hinders a real statistical evaluation, the results show that several of our processes under inert conditions facilitate the production of distinguishable flakes clearly below 100 nm in thickness. This comparison does not take into account the overall size and quality, *i.e.* monolithic and unbroken or aggregated structure, of the flakes, where TCNQ in THF easily exceeds the other systems, as it can be compared in the following Figure:

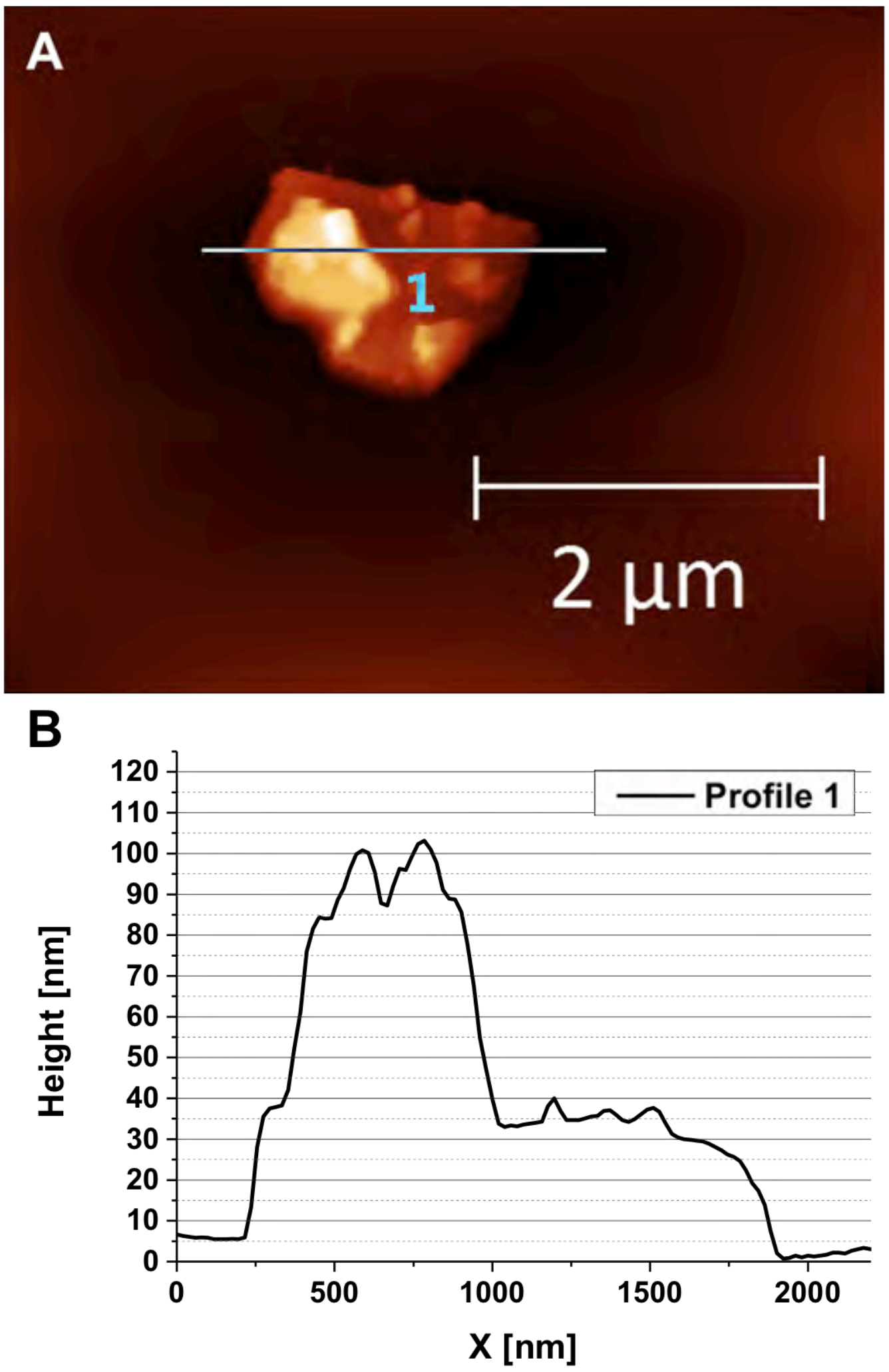

Large and thin BP flake, resulting from processing pristine BP in THF, assisted by TCNQ under inert atmosphere.

**Supplementary Figure 6 | Additional STEM-EELS data for pristine BP samples.**

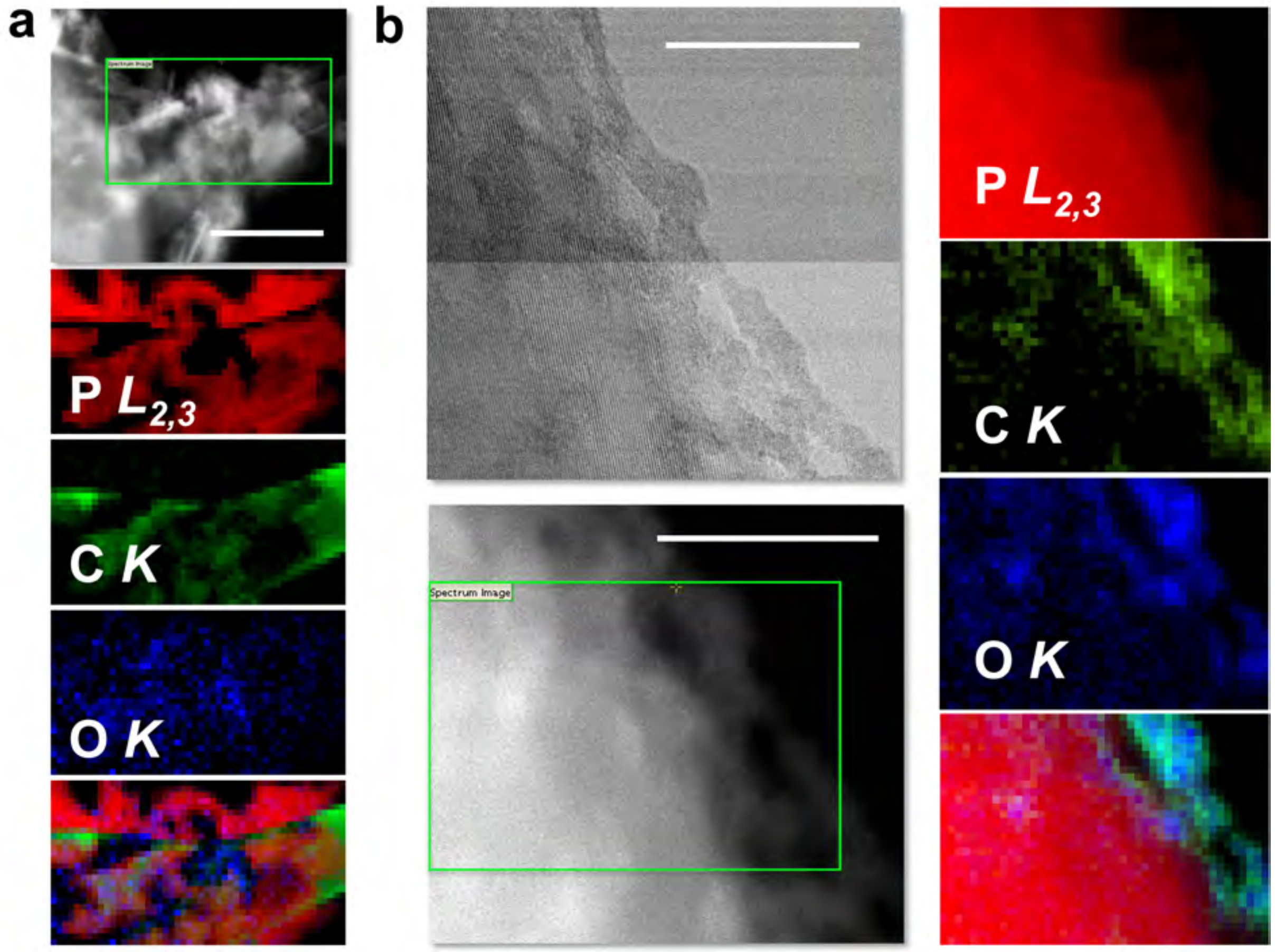

Pristine BP samples were also observed by aberration corrected STEM-EELS at 80 kV. (**a**) Low magnification images show that flakes are larger in size and more rugged and defective. The top HAADF image shows a pristine BP flake on a holey carbon support (visible in the carbon map below), with part of the flake hanging on the hole. The scale bar represents 500 nm. EELS mapping including the P $L_{2,3}$, C $K$, and O $K$ edges (red, green, and blue, respectively, from top to bottom) confirms that there is no carbon or oxygen on the flakes other than very minor contamination at the edges. The RGB image on the bottom right corner represents an overlay of the three compositional maps, identical color code. (**b**) ABF atomic resolution images (such as the one on the top left panel) show that the flakes are crystalline but edges are rugged and quite damaged. The maps on the right correspond to the P $L_{2,3}$, C $K$, and O $K$ edges (red, green, and blue, respectively, from top to bottom, along with RGB overlay). The HAADF image on the bottom left corner displays the region were this spectrum image was obtained, marked with a green rectangle. Very little carbon or oxygen signal is detected on the flake surface. The rugged edges may exhibit a very minor increase of C and O and occasionally other impurities ensuing from damage or contamination. The scale bars in (**b**) represent 20 nm. Principal component analysis was used to remove random noise. Compositional maps were obtained by acquiring and processing EEL spectrum images, the background was removed using power law fits and integrating the intensities remaining under the edges of interest (windows were 25 eV wide).

**Supplementary Figure 7 | Synthesis of the PDI derivative.**

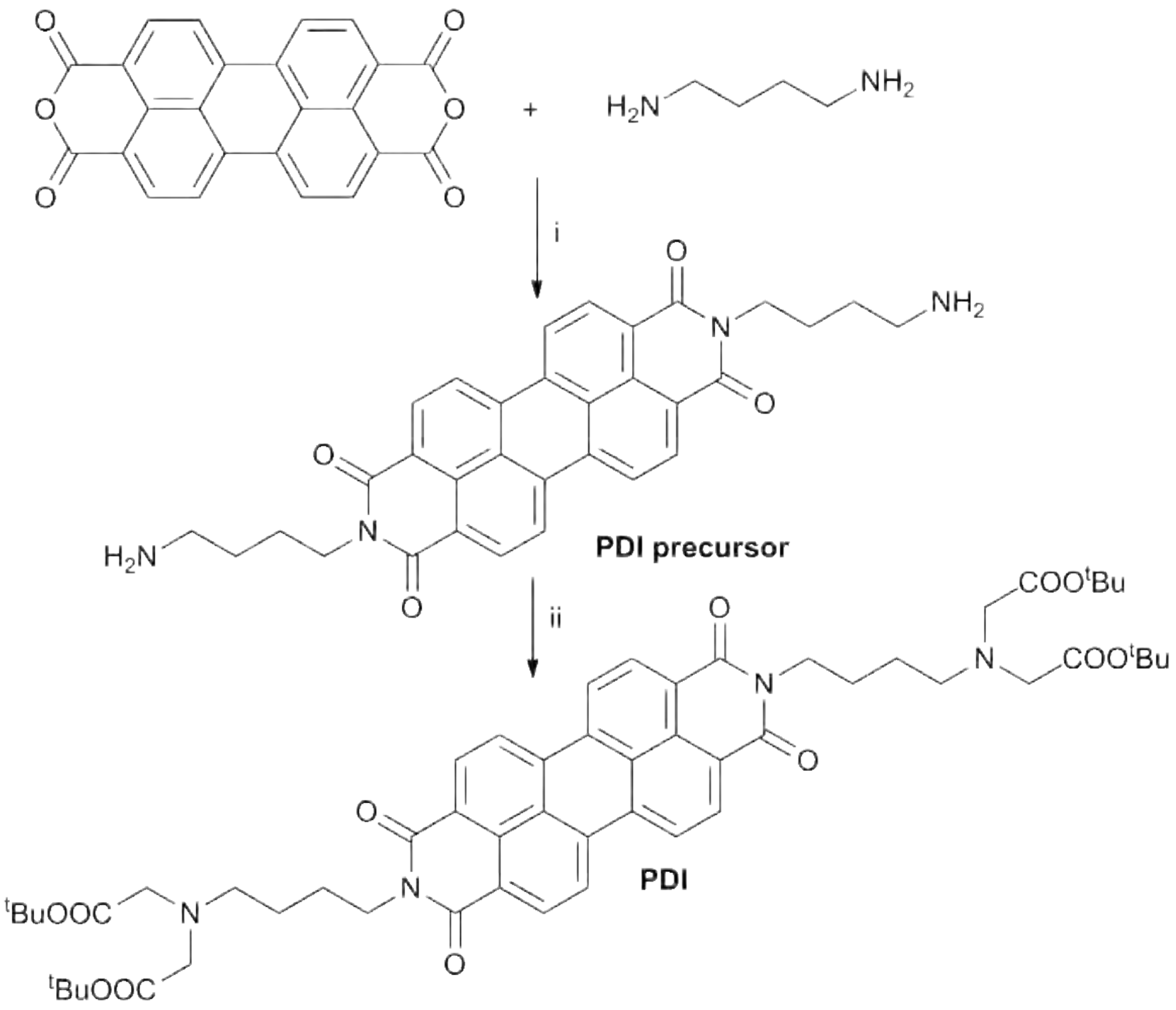

Scheme showing the synthesis of PDI molecule. i) Toluene, reflux, 4 h, yield: 89 %; ii) acetonitrile, diisopropylethylamine (DIPEA), *tert*-butyl bromoacetate, 60 °C, 24 h, yield: 13 %.

The PDI derivative has been synthesized according to Marcia *et al.*[19] Briefly, the PDI precursor was prepared by stirring a mixture of PTCDA (10 g, $2.6 \cdot 10^{-2}$ mol) and 1,4 diaminobutane (4 eq.) in toluene (250 mL) at reflux (110 °C) for 4 hours. After cooling down to room temperature, the mixture was filtered under *vacuum* and washed with toluene. The crude solid was then re-suspended in KOH 5 M (200 mL) and stirred for 15 hours at ambient temperature. Subsequently, the suspension was filtered and PDI precursor was collected as a red-brownish solid, which was dried in *vacuum* (16.4 g, yield = 89 %). Afterwards, a mixture of the PDI precursor (280 mg, $5.3 \cdot 10^{-4}$ mol), acetonitrile (15 mL), DIPEA (10 eq.), and *tert*-butyl bromoacetate (8 eq.) was stirred at 60 °C for 24 hours. Once cooled down to room temperature, it was *vacuum* filtered and the crude solid was washed with acetonitrile and water. Subsequently, the solid residue was dissolved in chloroform (5 mL) and hexane was added (100 mL). The mixture was stirred for 10 minutes at room temperature and then let stand for one night. The precipitate was filtered and dried under *vacuum*. The PDI is isolated as a brown solid (70 mg, yield = 13 %).

$^{1}$H NMR (300 MHz, $CDCl_3$, 25 °C): $\delta$ = 1.44 (s, 36H, 12 x $CH_3$), 1.63 (quintuplet, $J$ = 7.0 Hz, 4H, 2 x $CH_2$), 1.79 (quintuplet, $J$ = 7.5 Hz, 4H, 2 x $CH_2$), 2.77 (t, $J$ = 7.6 Hz, 4H, 2 x $CH_2$), 3.44 (s, 8H, 4 x $NCH_2$), 4.21 (t, $J$ = 7.4 Hz, 4H, 2 x $CH_2$), 8.33 (d, $J$ = 8.0 Hz, 4H, ArH), 8.48 (d, $J$ = 8.0 Hz, 4H, ArH) ppm.

$^{13}$C NMR (75 MHz, $CDCl_3$, 25 °C): $\delta$ = 25.54 (2 C, $CH_2$), 25.64 (2 C, $CH_2$), 28.05 (12 C, $CH_3$), 40.27 (2 C, $CH_2$), 53.86 (2 C, $CH_2$), 55.73 (4 C, $CH_2$), 80.73 (4 C, quat. C $^t$Bu), 122.47 (4 C, Ar-CH), 122.85 (2 C, Ar-C), 125.39 (2 C, Ar-C), 128.55 (4 C, Ar-C), 130.62 (4 C, Ar-CH), 133.58 (4 C, Ar-C), 162.76 (4 C, CON), 170.77 (4 C, COO) ppm.

MALDI-TOF (THAP): m/z 989 $(M+H)^+$, 1011 $(M + Na)^+$.

IR (ATR): $\nu$ = 2976.4, 2932.0, 1731.5, 1693.0, 1654.2, 1594.3, 1340.5, 1251.2, 1215.2, 1142.6, 988.1, 809.5, 745.9 $cm^{-1}$.

EA for $C_{56}H_{68}N_4O_{12}$: calcd. C 68.00, H 6.93, N 5.66; found C 67.66; H 6.90; N 5.65.

**Supplementary Figure 8 | Formation of charge-transfer complexes between PDI and NMP.**

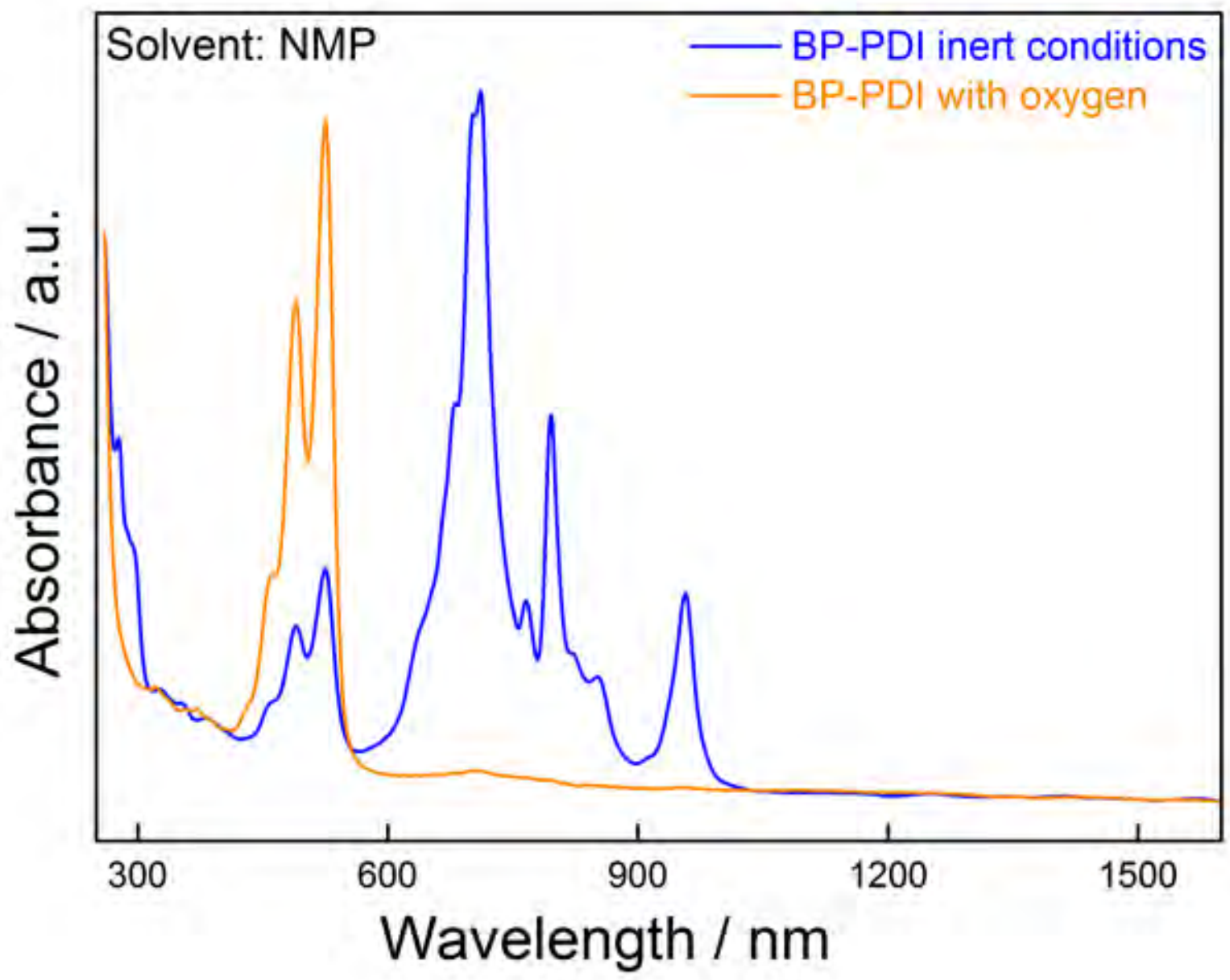

UV/Vis-nIR spectra showing the formation of the radical monoanion of the PDI by charge-transfer with the NMP under inert conditions. The radical monoanion species extremely sensitive and the neutral form can be recovered in presence of traces of oxygen.

In THF only the PDI is formed, whereas in NMP the $PDI^{\cdot-}$ appeared exhibiting and strong band around 700 nm corresponding to the transition between $D_0 \rightarrow D_n$, on the other hand, the lower energy band centered at *ca*. 956 nm can be assigned to the $D_0 \rightarrow D_1$ transition.[20]

**Supplementary Figure 9 | Optical-AFM-Raman relocalisation and correlation for BP–PDI.**

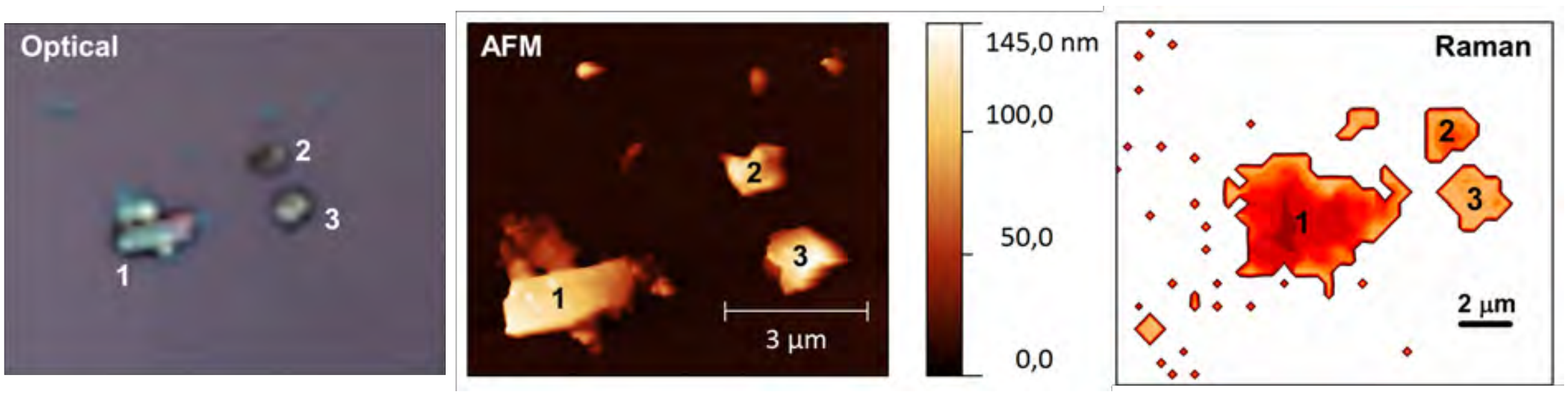

The relocalization makes use of the well-known optical contrast of nanomaterials deposited on opaque bilayers.[2,21] From left to right: optical micrograph (100x) of the selected area of study. Middle: AFM image of the same region identified following the procedure described in Supplementary Figure 3. Right: $A^1_g$ Raman map of the same region showing an excellent correlation.

**Supplementary Figure 10 | Quenching of the fluorescence of the PDI measured by Raman spectroscopy.**

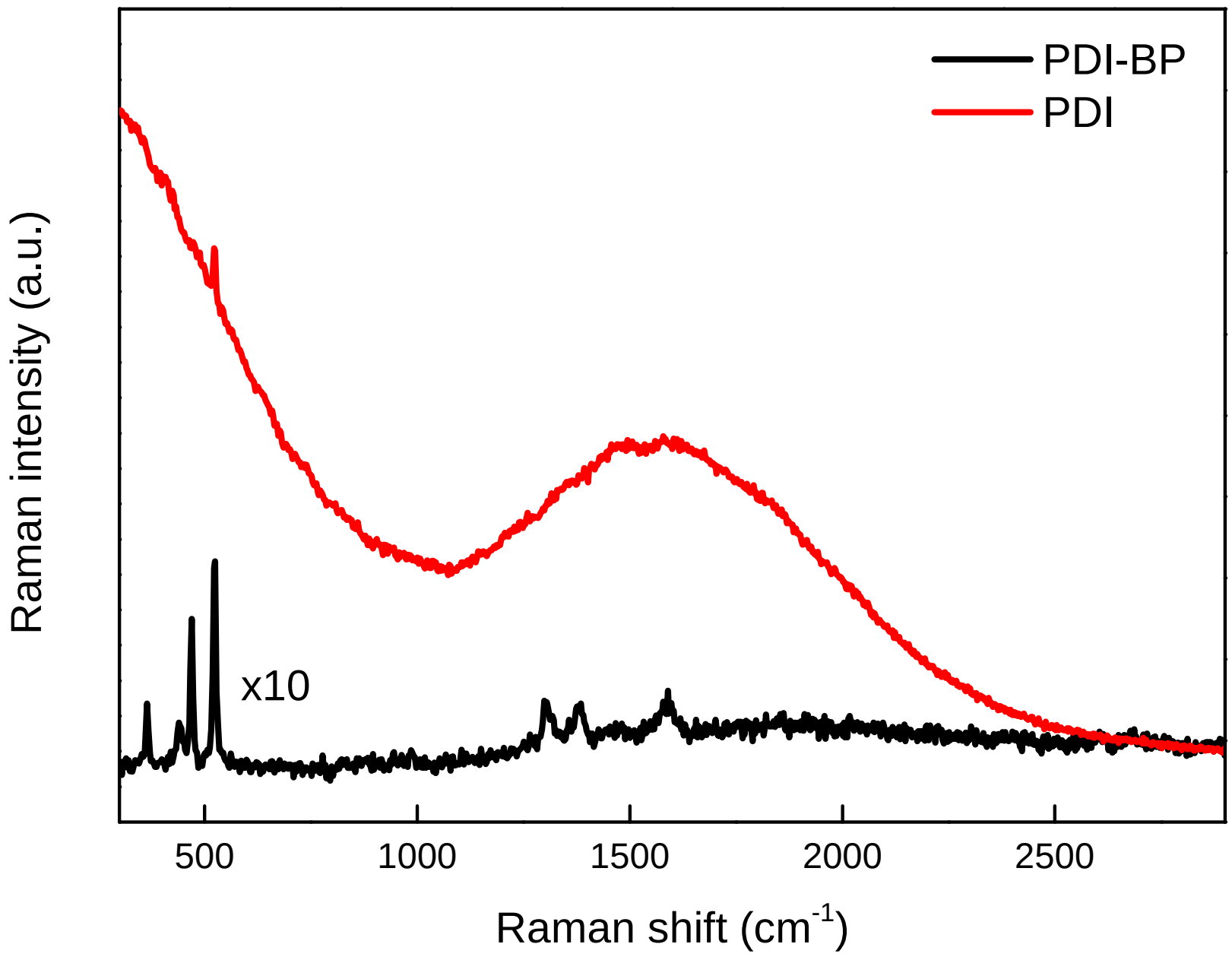

Raman spectra highlighting the quenching of the fluorescence of the PDI in the presence of BP. The spectrum of the PDI-BP has been magnified of a factor 10 for clarity. The characteristic signals of the PDI spectrum can be measured at: *ca*. 1302 and 1377 $cm^{-1}$ (in-plane ring "breathing"), 1457 $cm^{-1}$ (ring deformation), 1573 $cm^{-1}$ (in-plane C–C stretching), and 1700 $cm^{-1}$ (imide bending)

**Supplementary Figure 11 | Topographical AFM analysis of the BP–PDI sample.**

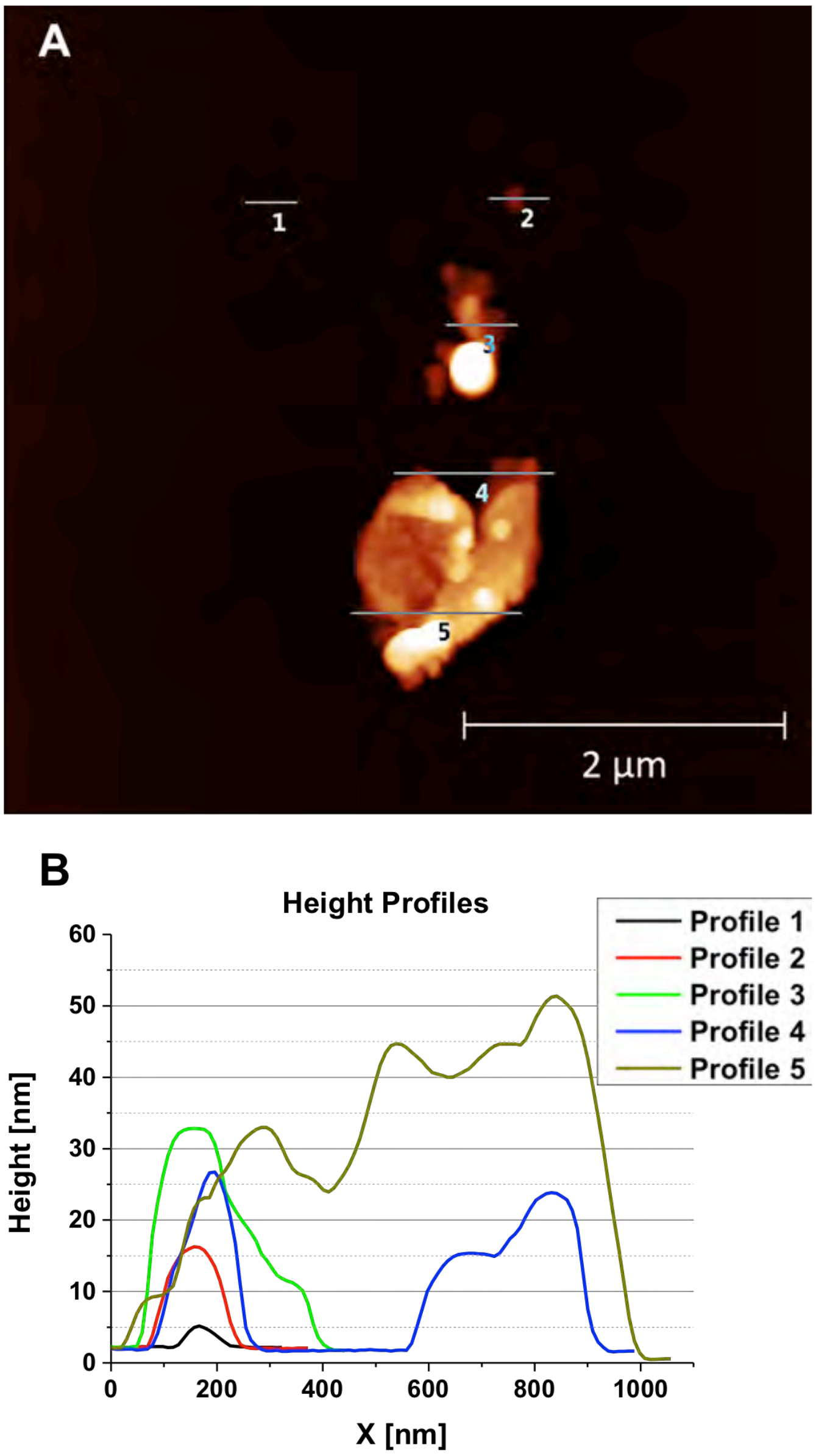

(**a**) AFM image showing the same region studied in Fig. 4 of the main text. (**b**) Topographic profiles of the areas highlighted in (**a**) showing the different height values of the sample.

**Supplementary Figure 12 | AFM topographic analysis of BP–PDI.**

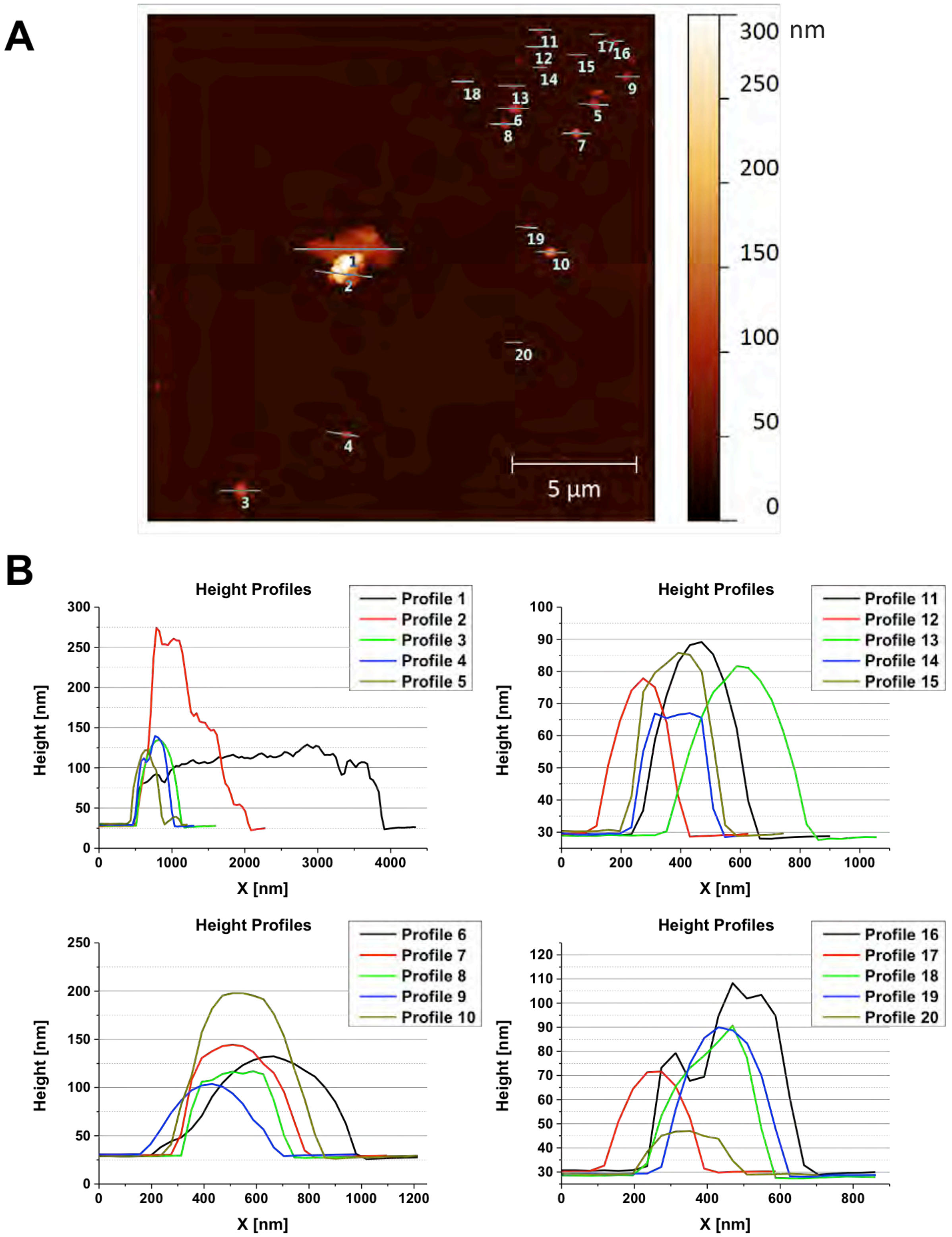

**Supplementary Figure 13 | Correlation between BP ($A^2_g$) and PDI ($\nu_{Ag}$) Raman intensities for BP–PDI.**

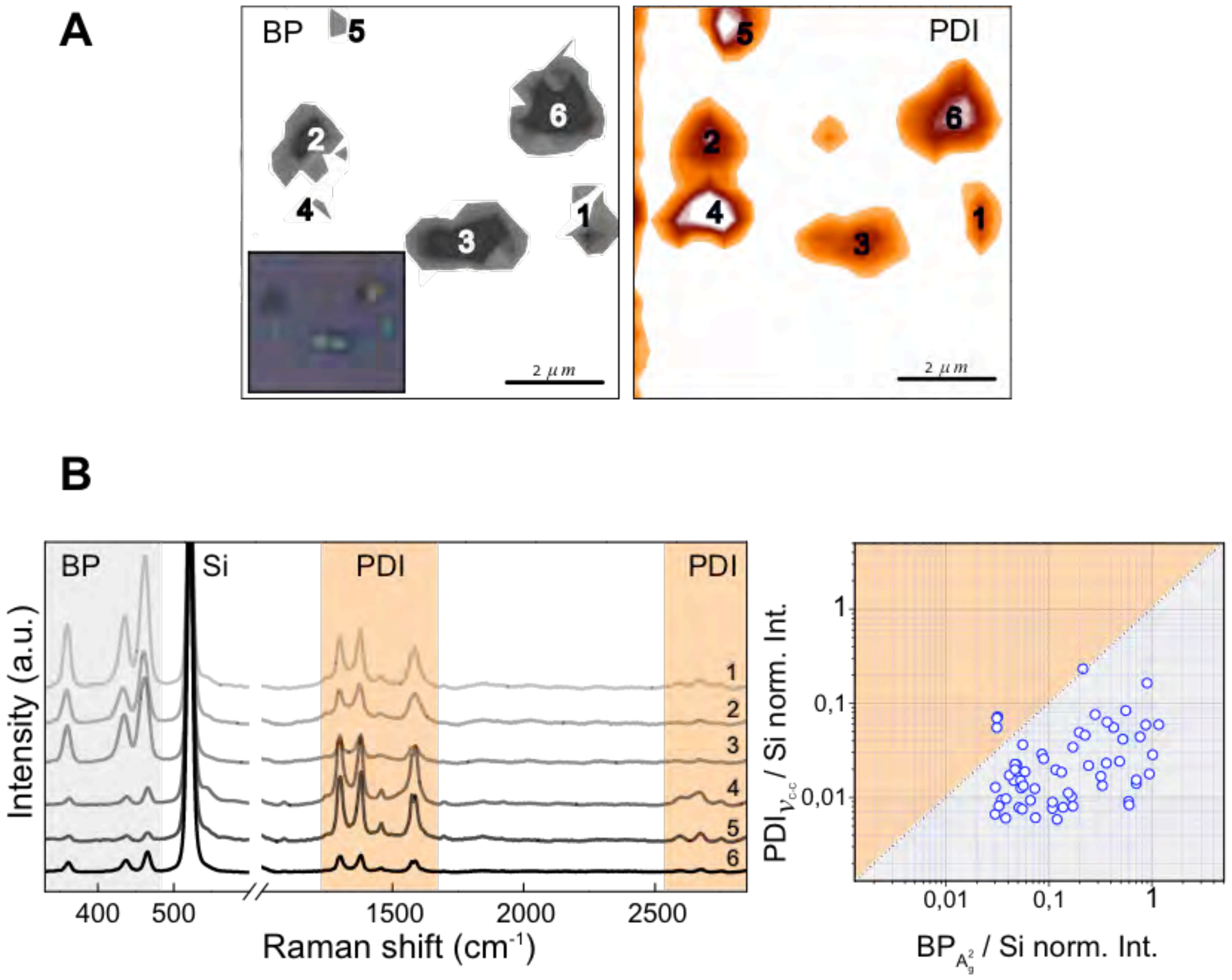

(**a**) Raman $A^2_g$ and $\nu_{Ag}$ intensity mappings of the same BP–PDI flakes deposited on a $Si/SiO_2$ substrate with a 300 nm thick oxide layer and shown in the inset (excitation wavelength 532 nm). The numbers denote the areas where the Raman spectra shown in (**b**) were taken. Interestingly, additional signals detected between 2500 $cm^{-1}$ and 2800 $cm^{-1}$, which originate from the side chains of the PDI, can be also observed. The main $A^1_g/A^2_g$-band ratio of each spectra exhibit the absence of intensity ratios <0.6, strongly suggesting that no oxidation has occurred.[2,16] (**b**, right) Plot of the $A^2_g$ versus $\nu_{Ag}$ normalized intensities (the normalization was developed using the Silicon band at 521 $cm^{-1}$) showing the relationship between the BP and the Raman enhancement effect, which is similar to that exhibited when $A^1_g$ is plotted instead of $A^2_g$.

**Supplementary Figure 14 | $A^1_g/A^2_g$ ratios of a BP–PDI sample after 48 h of exfoliation.**

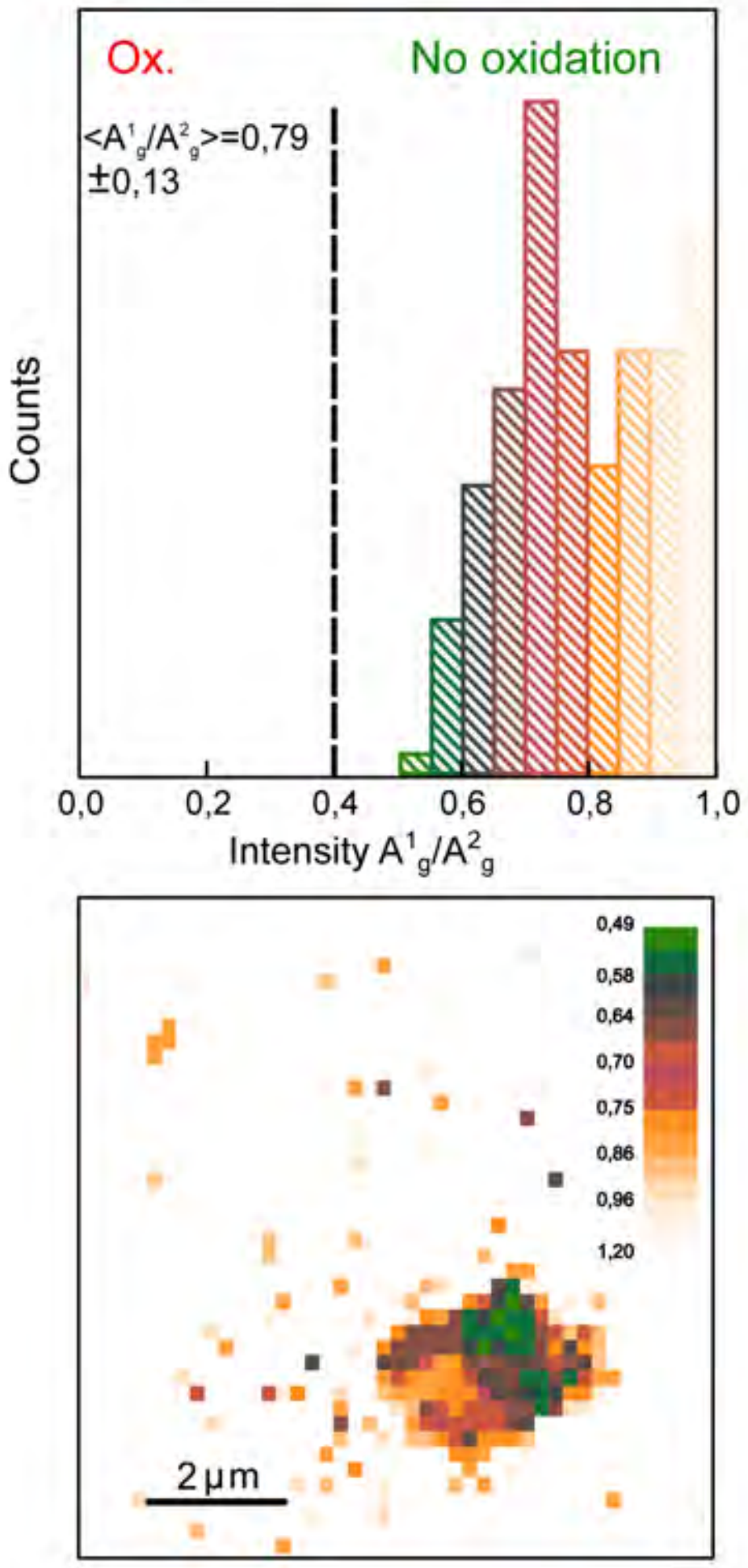

The Raman mapping of the $A^1_g/A^2_g$ ratio reveals an intact flake with no signals of oxidation, as reflected by the average $A^1_g/A^2_g$ ratio of 0.79.[16]

**Supplementary Figure 15: SRM study of a BP–PDI sample stored during 6 months.**

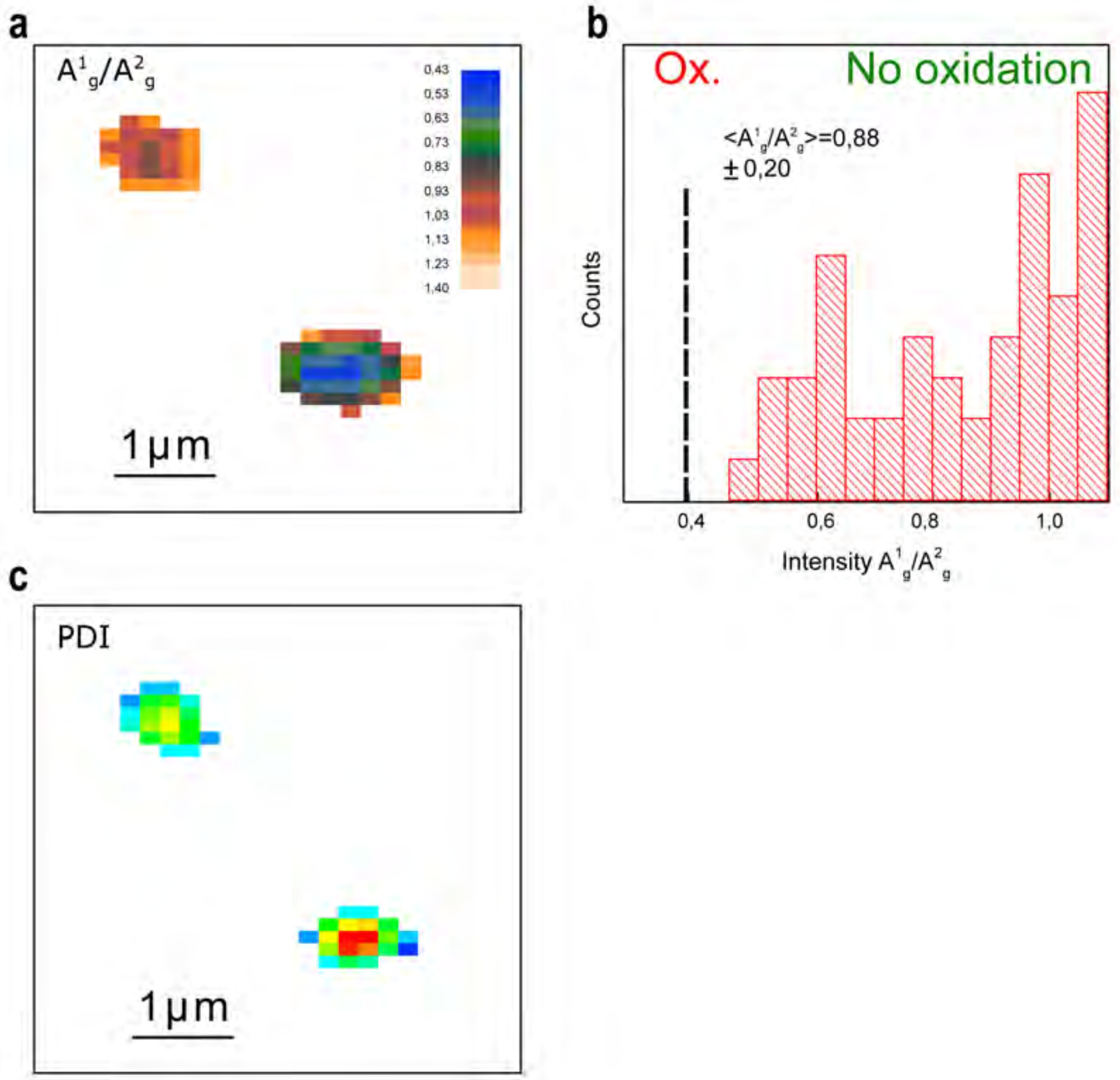

(**a**) The Raman mapping of the $A^1{}_g/A^2{}_g$ ratio reveals an intact flake with no signals of oxidation, as reflected by the average $A^1{}_g/A^2{}_g$ ratio of 0.88 (**b**).[16] (**c**) The corresponding $\nu_{Ag}$ Raman intensity mapping of the PDI.

**Supplementary Figure 16: Temperature-dependent Raman spectroscopy study of BP-PDI sample.**

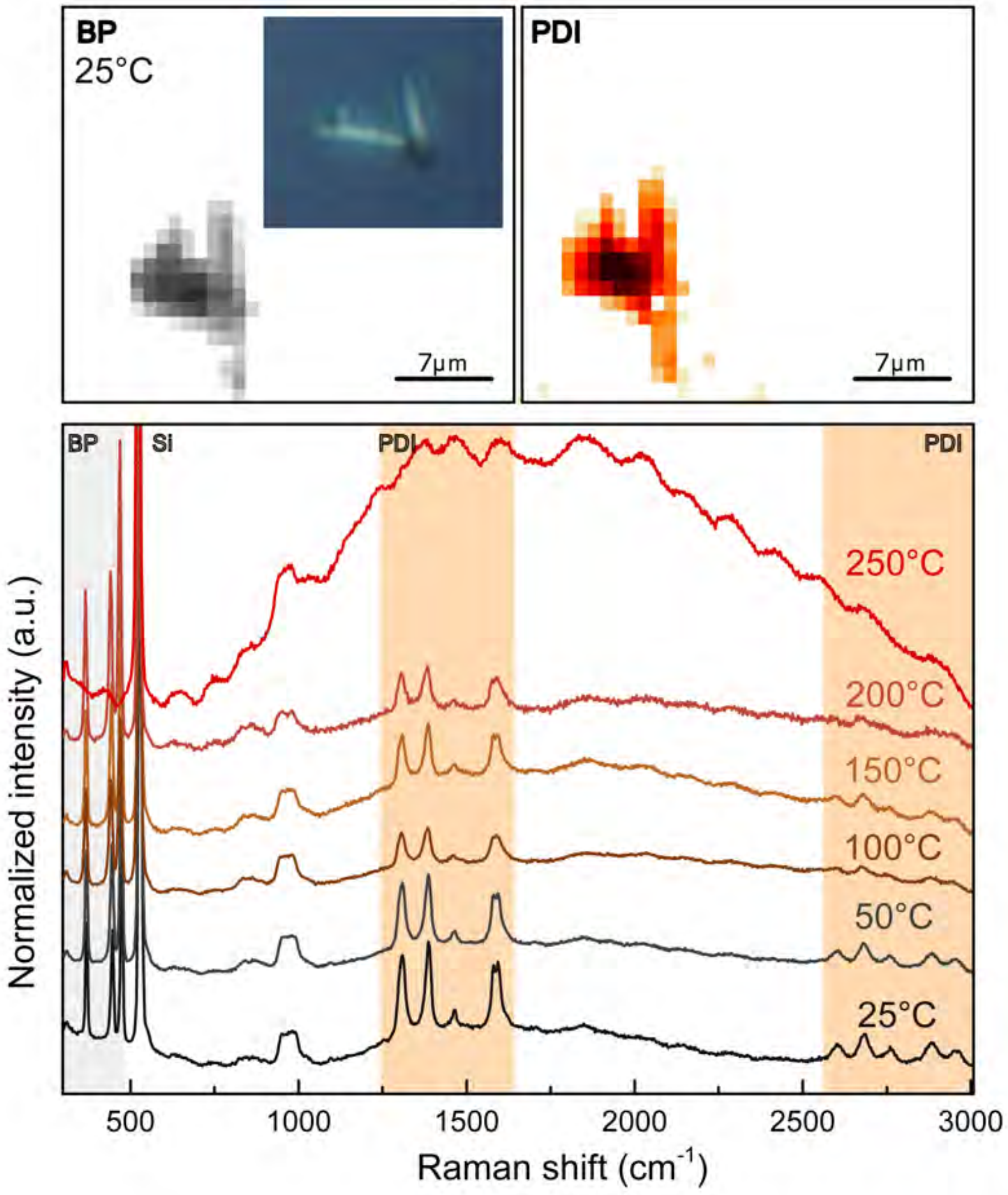

(**Top**) Raman $A_{1g}$ and $\nu_{Ag}$ intensity mappings obtained at room temperature in the temperature-controlled chamber. Optical micrograph of the same crystals is shown in the inset. (**Bottom**) Mean spectra acquired at different increasing temperatures ranging from 25 to 250 °C showing the stability of the sample below 200 °C. At higher temperatures, the characteristic peaks of the BP disappear as well as a strong photoluminescence from the PDI can be clearly seen (duration of the experiment *ca*. 2 hours).

**Supplementary Figure 17: Theoretical calculations.**

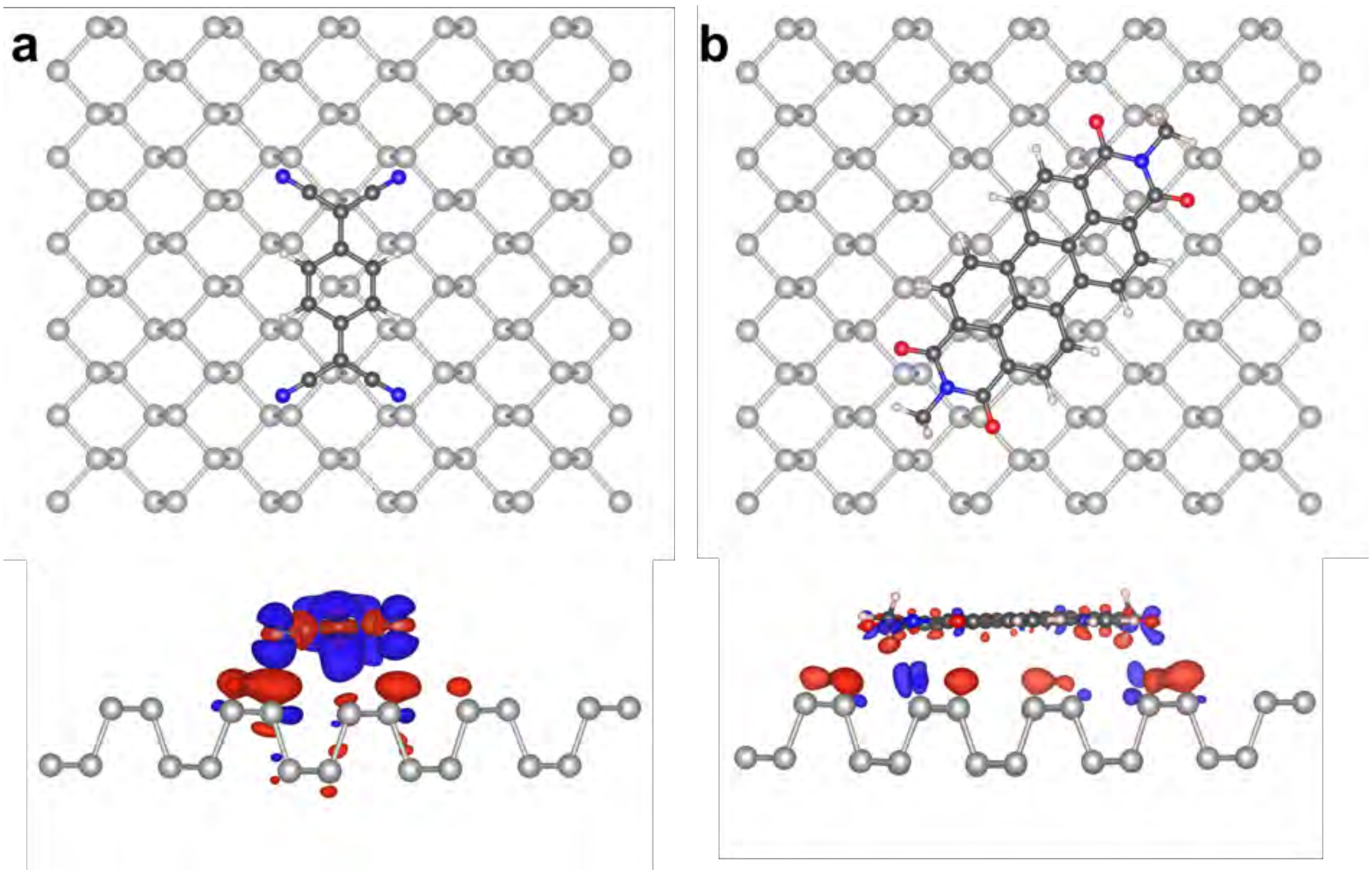

**Structural models for TCNQ and PDI on single-layer BP.** (**a**) TCNQ on SL-BP, top and side view. (**b**) PDI on SL-BP, top and side view. In both cases the electron density difference between the BP-TCNQ/BP-PDI and the isolated molecules and BP is also shown in the side view (for an isovalue of $\pm 0.0005|e|/Å^3$, blue/red indicates higher/lower electron density, respectively).

To shed light on the nature of the interactions between the two studied molecules TCNQ and PDI, we also performed DFT calculations. First, the interaction of TCNQ and PDI with single-layer BP (SL-BP) was investigated. In case of TCNQ we considered different adsorption geometries on the surface of SL-BP but found that the results did not depend significantly on the adsorption geometry. The most stable configuration is depicted in Fig. S17a, which is the same structure as found by Jing *et al.* in a recent study.[22] The adsorption energy of TCNQ on SL-BP is found to be between −1.31 to –1.29 eV (depending on the relative orientation between TCNQ and the BP monolayer), which indicates a rather strong non-covalent interaction between TCNQ and SL-BP. (We define adsorption energies per molecule as: $E_a = E_{BP/molecule} - E_{BP} - E_{molecule}$, where $E_{BP/molecule}$, $E_{BP}$ and $E_{molecule}$ stand for the total energy of the molecule-modified BP, the isolated BP and the isolated molecule, respectively.) The analysis of the energy contributions reveals that by far the most of the binding energy (98%) is contributed from the van-der-Waals correction, indicating that TCNQ is mostly bound by van-der-Waals interactions. The distance between TCNQ and SL-BP is 3.1 Å. Due to this short distance and since TCNQ is a strong electron acceptor, we observe a charge transfer of 0.43 |e| from SL-BP to TCNQ as calculated by a Bader population analysis. This can be also seen from the electron density difference between BP-TCNQ and the separated molecules, see Fig S17a (lower panel). As explained in the preceding sections both IR/Raman and UV/Vis spectra indicate that TCNQ is negatively charged on BP, with especially the UV/Vis spectra showing an intriguingly high similarity to known $TCNQ^{2-}$ compounds while there is no indication for $TCNQ^{\cdot-}$. From a naïve point of view one could expect that along with increasing charge transfer from BP to TCNQ first the $TCNQ^{\cdot-}$ and then the $TCNQ^{2-}$ spectra should emerge. However, due to our calculations the BP-TCNQ system is closed shell, i.e. exhibits no radical character. Thus, the typical spectrum of $TCNQ^{\cdot-}$ will not occur, but the UV/Vis spectrum would change continuously from the neutral TCNQ directly to $TCNQ^{2-}$ with increasing charge on TCNQ placed on BP. This makes it plausible why already a charge transfer smaller than 2 |e| could yield an UV/Vis spectrum similar to that of $TCNQ^{2-}$. Finally, vibrational frequencies of TCNQ and TCNQ on SL-BP (ignoring solvent effects) were calculated. We find that already in our model the CN-bands get downshifted by 20–25 $cm^{-1}$, similar is true for the modes $\nu_{20}$, $\nu_{34}$, and $\nu_{50}$ (see supplementary Table 1), in agreement with the experimental data.

The adsorption energy of PDI is –2.01 eV, but in contrast to TCNQ the binding is purely due to van-der-Waals interactions, which is illustrated by the fact that the contribution from the van-der-Waals correction to the binding energy is -2.30 eV, i.e. without this correction PDI would be repelled from the BP monolayer. Moreover, the charge transfer from SL-BP to PDI is much more limited, 0.10 |e|, although the distance between PDI and SL-BP (3.0 Å) is similar to that of TCNQ and SL-BP.

In case of TCNQ the adsorption on double-layer BP (DL-BP) and intercalation between two monolayers was investigated as well. The binding energy of TCNQ on DL-BP is even less dependent on the orientation than on SL-BP and amounts to –1.58 eV, i.e. is more stable than on SL-BP; the charge transfer increases to 0.65 |e|. Intercalation of TCNQ between two BP monolayers is thermodynamically not stable, but the charge transfer is then even more pronounced: 0.97 |e|. This shows that TCNQ can get highly charged by BP by noncovalent interaction with BP.

**Supplementary Table 1. Calculated vibrational frequencies for TCNQ and SL-BP-TCNQ.**

| mode | description | TCNQ | SL-BP-TCNQ |
|---|---|---|---|
| $\nu_{33}$ | C≡N stretch | 2257 | 2238 |
| $\nu_{1g}$ | C≡N stretch | 2237 (very weak) | 2211 |
| $\nu_{20}$ | C=C stretch | 1531 | 1497 |
| $\nu_{34}$ | C=C stretch | 1526 | 1508 |
| $\nu_{50}$ | C-H out of plane | 849 | 832 |

Selected vibrational frequencies of TCNQ and TCNQ on single-layer BP as calculated by finite differences, values in $cm^{-1}$. The numbering of the modes is as in Ref. 20 of the main paper. Solvent effects were ignored.

**Supplementary Figure 18: Influence of the step size and the oxidation degree in the SRM mappings of micromechanically-exfoliated flakes.**

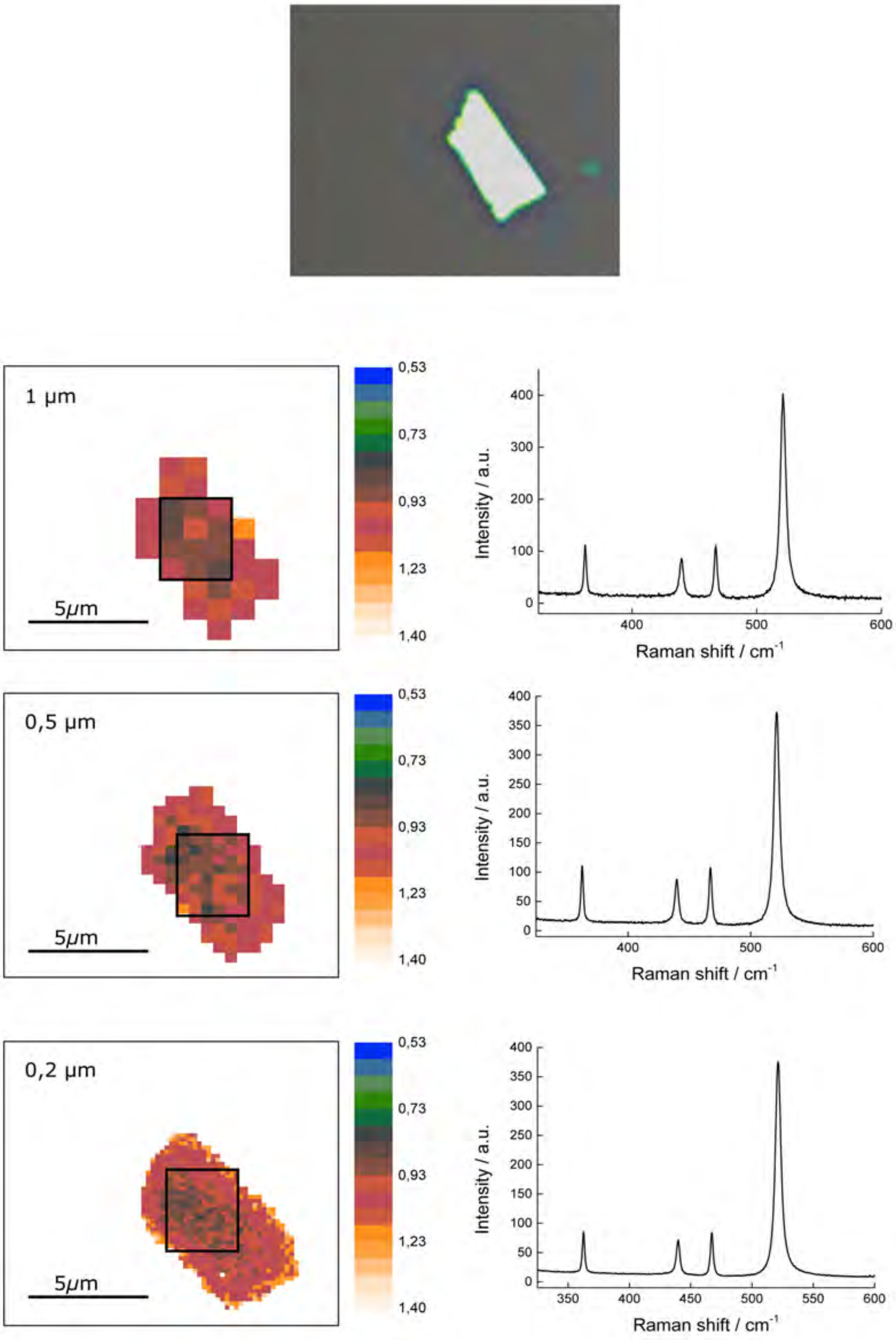

Influence of the step size in the SRM mappings. Optical micrograph of the micromechanically exfoliated flake under study. The exfoliation was developed in the glove box under strictly inert conditions. After the exfoliation, the flake was transferred to a Si/$SiO_2$ wafer. The SRM was performed under inert conditions to exclude any influence of the oxidation on the Raman data. (**Top**) One-micron spot size and its corresponding mean average spectra. (**Middle**) 0.5 microns spot size. (**Bottom**) 0.2 microns spot size. All the average spectra exhibited exactly the same $A_{1g}/A_{2g}$ ratio.

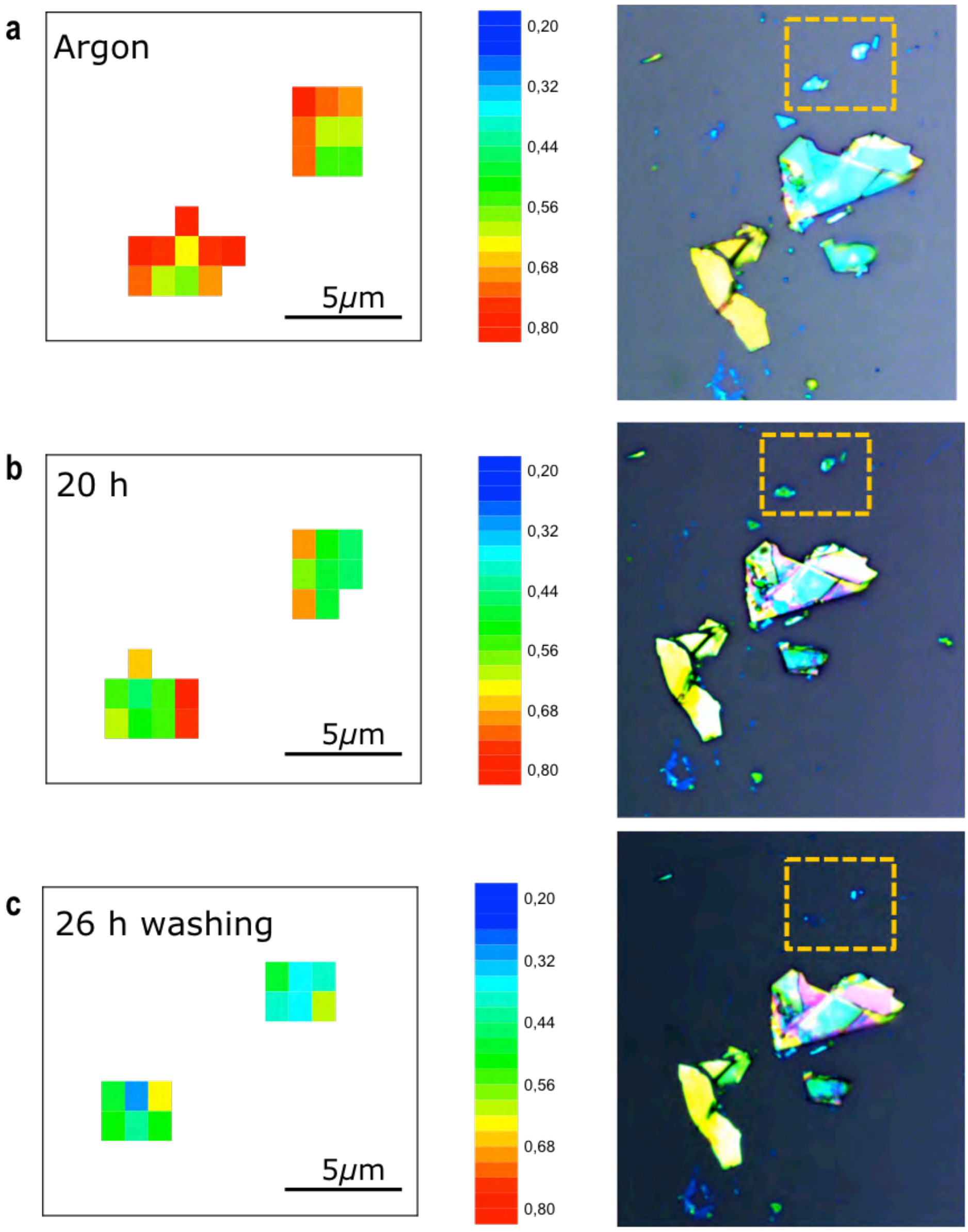

SRM study of the influence of the oxidation degree on micromechanically exfoliated flakes measured first under inert conditions (**a**), after 20 h of exposure to the ambient (**b**), and finally after washing the flakes with water (**c**). Interestingly, the $A_{1g}/A_{2g}$ ratio decreased with the oxidation degree showing final values below the 0.4 limit, thus indicating a complete oxidation of the flakes.[16] Moreover, the degradation of the black phosphorus is also reflected in the decrease of the lateral dimensions. This is evident from both the SRM mappings and the optical micrographs. The step size was 1 micron.

## Supplementary References.

1. Kang, J. *et al.* Solvent Exfoliation of Electronic-Grade, Two-Dimensional Black Phosphorus. *ACS Nano* **9,** 3596–3604 (2015).

2. Hanlon, D. *et al.* Liquid exfoliation of solvent-stabilized few-layer black phosphorus for applications beyond electronics. *Nat. Commun.* **6,** 8563 (2015).

3. Kresse, G. & Furthmüller, J. Efficiency of ab-initio total energy calculations for metals and semiconductors using a plane-wave basis set. *Comput. Mater. Sci.* **6,** 15–50 (1996).

4. Kresse, G. & Furthmüller, J. Efficient iterative schemes for \textit{ab initio} total-energy calculations using a plane-wave basis set. *Phys. Rev. B* **54,** 11169–11186 (1996).

5. Perdew, J. P., Burke, K. & Ernzerhof, M. Generalized Gradient Approximation Made Simple. *Phys. Rev. Lett.* **77,** 3865–3868 (1996).

6. Grimme, S., Antony, J., Ehrlich, S. & Krieg, H. A consistent and accurate ab initio parametrization of density functional dispersion correction (DFT-D) for the 94 elements H-Pu. *J. Chem. Phys.* **132,** 154104 (2010).

7. Grimme, S., Ehrlich, S. & Goerigk, L. Effect of the damping function in dispersion corrected density functional theory. *J. Comput. Chem.* **32,** 1456–1465 (2011).

8. Bučko, T., Hafner, J. & Ángyán, J. G. Geometry optimization of periodic systems using internal coordinates. *J. Chem. Phys.* **122,** 124508 (2005).

9. Monkhorst, H. J. & Pack, J. D. Special points for Brillouin-zone integrations. *Phys. Rev. B* **13,** 5188–5192 (1976).

10. Blöchl, P. E., Jepsen, O. & Andersen, O. K. Improved tetrahedron method for Brillouin-zone integrations. *Phys. Rev. B* **49,** 16223–16233 (1994).

11. Methfessel, M. & Paxton, A. T. High-precision sampling for Brillouin-zone integration in metals. *Phys. Rev. B* **40,** 3616–3621 (1989).

12. Khatkale, M. S. & Devlin, J. P. The vibrational and electronic spectra of the mono-, di-, and trianon salts of TCNQ. *J. Chem. Phys.* **70,** 1851–1859 (1979).

13. Suchanski, M. R. & Van Duyne, R. P. Resonance Raman spectroelectrochemistry. IV. The oxygen decay chemistry of the tetracyanoquinodimethane dianion. *J. Am. Chem. Soc.* **98,** 250–252 (1976).

14. Van Duyne, R. P. *et al.* Resonance Raman spectroelectrochemistry. 6. Ultraviolet laser excitation of the tetracyanoquinodimethane dianion. *J. Am. Chem. Soc.* **101,** 2832–2837 (1979).

15. Kamitsos, E. I. & Jr, W. M. R. Raman studies in CuTCNQ: Resonance Raman spectral

observations and calculations for TCNQ ion radicals. *J. Chem. Phys.* **79,** 5808–5819 (1983).

16. Favron, A. *et al.* Photooxidation and quantum confinement effects in exfoliated black phosphorus. *Nat. Mater.* **14,** 826–832 (2015).

17. Yasaei, P. *et al.* High-Quality Black Phosphorus Atomic Layers by Liquid-Phase Exfoliation. *Adv. Mater.* **27,** 1887–1892 (2015).

18. Woomer, A. H. *et al.* Phosphorene: Synthesis, Scale-Up, and Quantitative Optical Spectroscopy. *ACS Nano* **9,** 8869–8884 (2015).

19. Marcia, M., Singh, P., Hauke, F., Maggini, M. & Hirsch, A. Novel EDTA-ligands containing an integral perylene bisimide (PBI) core as an optical reporter unit. *Org. Biomol. Chem.* **12,** 7045 (2014).

20. Gosztola, D., Niemczyk, M. P., Svec, W., Lukas, A. S. & Wasielewski, M. R. Excited Doublet States of Electrochemically Generated Aromatic Imide and Diimide Radical Anions. *J. Phys. Chem. A* **104,** 6545–6551 (2000).

21. Backes, C., Englert, J. M., Bernhard, N., Hauke, F. & Hirsch, A. Optical Visualization of Carbon Nanotubes—a Unifying Linkage Between Microscopic and Spectroscopic Characterization Techniques. *Small* **6,** 1968–1973 (2010).

22. Jing, Y., Tang, Q., He, P., Zhou, Z. & Shen, P. Small molecules make big differences: molecular doping effects on electronic and optical properties of phosphorene. *Nanotechnology* **26,** 095201 (2015).